\documentclass[notitlepage,superscriptaddress]{revtex4-1}
\usepackage{tikz}
    \usetikzlibrary{shapes.geometric,arrows}

    \tikzset{
        tree node/.style={
            draw
        },
        tree edge/.style={
            ->,
            thick,
            >=stealth
        },
        model selection node/.style={draw=white, 
            fill=cud-reddish-purple, 
            circle,
            minimum size = 0.6cm,
            text=white, scale=1.5
        },
        mixture model node/.style={draw=white,
            fill=cud-sky-blue,
            regular polygon, regular polygon sides=3,
            minimum size = 0.8cm,
            text=white, scale=1.5
        },
        particle filter node/.style={draw=white, 
            fill=cud-vermillion, 
            minimum size = 0.6cm,
            text=white, scale=1.5
        }
    }

\usepackage[pretty]{revquantum}
\usepackage{algorithm}
\usepackage{algpseudocode}

    \renewcommand{\inlinecomment}[1]{\Comment {\footnotesize #1} \normalsize}
    \renewcommand{\linecomment}[1]{\State \(\triangleright\) {\footnotesize #1} \normalsize}
    \newcommand{\seccomment}[1]{%
        \vskip0.5em%
        \State {\color{gray} \footnotesize \(\blacksquare\) \emph{#1}} \normalsize
    }

    \newrm{iter}
    \newrm{iters}
    \newrm{cluster}
    \newrm{ess}
    \newcommand{\bfop}{\operatorname{bf}}
    \newcommand{\argmin}{\operatorname{arg min}}
    \newcommand{\minnpart}{\ensuremath{n_{\min,\textrm{part}}}}
    \newcommand{\npart}{\ensuremath{n_{\textrm{part}}}}
    \newcommand{\nexp}{\ensuremath{n_{\textrm{exp}}}}
    \newcommand{\iexp}{\ensuremath{i_{\textrm{exp}}}}
    \algnewcommand\Raise{\textbf{raise} }%
    \newcommand{\kof}[4][\in]{\ensuremath{%
        \left\{%
            #2_{#3} : #3 #1 \{1, \dots, #4\}%
        \right\}%
    }}

\usepackage[bold]{hhtensor}
\usepackage{complexity}

\usepackage[caption=false]{subfig}

\newaffil{MSRQuArC}{
    Station Q Quantum Architectures and Computation Group,
	Microsoft Research,
    Redmond, WA, United States
}

\newcommand{\figurefolder}{../fig}
\newcommand{\eq}[1]{Eq.~\hyperref[eq:#1]{(\ref*{eq:#1})}}
\renewcommand{\sec}[1]{\hyperref[sec:#1]{Section~\ref*{sec:#1}}}
\newcommand{\app}[1]{\hyperref[app:#1]{Appendix~\ref*{app:#1}}}
\newcommand{\apx}[1]{\hyperref[apx:#1]{Appendix~\ref*{apx:#1}}}
\newcommand{\tab}[1]{\hyperref[tab:#1]{Table~\ref*{tab:#1}}}
\newcommand{\fig}[1]{\hyperref[fig:#1]{Figure~\ref*{fig:#1}}}
\newcommand{\figa}[2]{\hyperref[fig:#1]{Figure~\ref*{fig:#1}#2}}
\newcommand{\figx}[2]{\hyperref[fig:#1]{Figure~\ref*{fig:#1}(#2)}}
\newcommand{\thm}[1]{\hyperref[thm:#1]{Theorem~\ref*{thm:#1}}}
\newcommand{\lem}[1]{\hyperref[lem:#1]{Lemma~\ref*{lem:#1}}}
\newcommand{\cor}[1]{\hyperref[cor:#1]{Corollary~\ref*{cor:#1}}}
\newcommand{\defn}[1]{\hyperref[def:#1]{Definition~\ref*{def:#1}}}
\newcommand{\alg}[1]{\hyperref[alg:#1]{Algorithm~\ref*{alg:#1}}}
\newcommand{\prob}[1]{\hyperref[prob:#1]{Problem~\ref*{prob:#1}}}

\usepackage{array,booktabs,tabularx}
\newcolumntype{C}{>{$}c<{$}}
\AtBeginDocument{
    \heavyrulewidth=.08em
    \lightrulewidth=.05em
    \cmidrulewidth=.03em
    \belowrulesep=.65ex
    \belowbottomsep=0pt
    \aboverulesep=.4ex
    \abovetopsep=0pt
    \cmidrulesep=\doublerulesep
    \cmidrulekern=.5em
    \defaultaddspace=.5em
}
\newcommand{\parametertable}[1]{
    \vskip1.2em
    \begin{tabularx}{0.6\linewidth}{lXlX}
        \toprule
        \multicolumn{4}{l}{\emph{Parameters}} \\
        \multicolumn{2}{l}{Common} & \multicolumn{2}{l}{Structured} \\
        \hline
        #1
        \bottomrule
    \end{tabularx}
}

\renewcommand{\figurefolder}{fig}
\begin{document}

\title{Structured Filtering}

\author{Christopher Granade}
\affilUSydPhys
\affilEQuSUSyd

\author{Nathan Wiebe}
\affilMSRQuArC

\begin{abstract}
A major challenge facing existing sequential Monte-Carlo methods for parameter estimation in physics stems from the inability of
existing approaches to robustly deal with experiments that have different mechanisms that yield the results with equivalent probability.
We address this problem here by proposing a form of particle filtering that clusters the particles that comprise the sequential Monte-Carlo
approximation to the posterior before applying a resampler.  Through a new graphical approach to thinking about such models, we are
able to devise an artificial-intelligence based strategy that automatically learns the shape and number of the clusters in the support of 
the posterior.  We demonstrate the power of our approach by applying it to randomized gap estimation and a form of low circuit-depth phase
estimation where existing methods from the physics literature either exhibit much worse performance or even fail completely.
\end{abstract}

\maketitle

\section{Introduction}
\label{sec:intro}

Across a range of physical sciences, much of the work of the experimentalist centers around learning from observed data. In metrology, for instance, experimental data
is used to infer parameters of interest such as  magnetic
fields, temperature, or other physical quantities
\cite{giovannetti_advances_2011}. The process
by which these parameters are learned from data is thus of critical
importance to tasks as diverse as metrology
and quantum information
processing~\cite{berry2009perform,blume2010optimal,huszar2012adaptive,granade_robust_2012,stenberg2014efficient}.

The importance of learning physical parameters has motivated developing
and making practical advances in statistically-principled approaches to
parameter estimation. Bayesian methods in general and Bayesian inference
in particular have proven to provide a compelling framework for drawing
inferences from experimental observations in a rigorous, robust and practical
manner~\cite{mackay2003information}. Numerical algorithms such as particle filtering then
provide a general and practical framework for approximate Bayesian inference, as well
as for statistical computation more generally \cite{doucet_tutorial_2011}.

In practice, existing approaches suffer from a great deficiency: they implicitly assume that
the probability distribution that describes the current state of knowledge about the parameters
has a particular structure to it.  These assumptions are often reasonable, such as the assumption
that the probability distribution is Gaussian or, weaker still, unimodal.  While these reasonable sounding
assumptions often work well, they can fail in catastrophic ways when evidence is provided
that favors a multitude of equivalent (or near-equivalent) hypotheses.  We refer to such
learning problems as \emph{degenerate} in analogy to quantum mechanics.  The challenges posed by
degeneracy can sometimes be circumvented through the use
of cunningly designed experiments, but this places additional experimental
demands, and is not always possible.  As a result, there are broad
classes of parameter estimation problems for which we have no robust automated methods for parameter estimation.

In this paper, we allowing the inference algorithm to learn the structure of its posterior distribution over the parameters of interest.
We show that doing so also allows for relaxing these experimental constraints.
Specifically, we demonstrate that this can allow such algorithms to learn when existing methods fail. Our algorithm augments traditional particle filtering methods with
a dynamically generated tree describing the structure of a Bayesian posterior
as a hierarchal clustering of particle filters. We rely on statistically
principled approaches to model selection to remove structural elements that
are inconsistent with the observed data, ensuring that the trees generated by
our algorithm usefully represent its state of knowledge.

The advantages offered by our structured filtering algorithm are especially
relevant in the case of degeneracy, which presents a significant
challenge to existing work. Conventional methods are either inefficient or
subtly bias the inference towards unimodal distributions. If the model for a
learning problem contains two sets of parameters that are nearly equally
likely for the data observed, these efficient inference algorithms usually
fail in catastrophic ways. Methods for dealing with this, such as annealing~\cite{del2006sequential}, qualitatively
fail to give us a solution because they cause the solution to focus on just
one of the families of degenerate models.  Commonly used solutions such as these are therefore, at best, imperfect solutions.  Furthermore, these limitations
prevent the application of these methods to cases where dynamical systems are
probed using uninformative experiments, which often yield outcomes that are
consistent with several hypothetical dynamical models; that in turn places
several experimental constraints that are relaxed by our algorithm.

We begin in \autoref{sec:bayes} by reviewing Bayesian inference as a 
framework for learning parameters from experimental observations, then introduce
our structured filtering algorithm for Bayesian inference in \autoref{sec:structured}.
In \autoref{sec:results} and \autoref{sec:phase-est}, we present numerical evidence for
the efficiency and correctness of our approach before concluding in \autoref{sec:conclusion}.

\section{Bayesian Inference}
\label{sec:bayes}

Bayesian inference has become a fundamental tool for modeling quantum systems.  The basic object in Bayesian inference is the prior distribution.  The prior distribution is a probability distribution over a set of hypotheses that could describe the system.  In particular, let us assume that we have a model for a data set that is parameterized by $\vec{x}$ then the prior distribution is $P(\vec{x})$.  The prior distribution represents any beliefs that the experimenter may hold about the true model before processing any subsequent data.  While the prior distribution is subjective, the Bernstein-von Mises theorem~\cite{freedman_wald_1999} states that under most circumstances a poorly chosen prior will simply slow down the inference.  This subjectivity has made the use of prior distributions a source of contention between frequentists and Bayesian statisticians.  If one dislikes the Bayesian interpretation of the prior it is possible to eschew the discussion of the prior (almost) completely by making the initial prior uniform and avoiding adaptive strategies.  Moreover, by performing Bayesian inference on an artificial likelihood function,
one can derive a useful approximation of maximum-likelihood estimation~\cite{mackay2003information}, such that the utility
of Bayesian inference does not hinge on adopting a particular philosophical view.
That said, we will take a Bayesian interpretation of probability in the following as a matter of
convenience.

The likelihood function is the second component of Bayesian inference.  Its purpose is to compute the probability of observing a vector of experimental outcomes ${\bf D}$ given that a hypothesis $\vec{x}$ is true.
It is denoted $P({\bf D}|\vec{x}; t)$.  For example, in quantum mechanics consider $\vec{x} = [\omega]$ to be a Rabi frequency for the Hamiltonian
$H(\vec{x}) = \omega \sigma_x/2$. Then given a state $e^{-i H(\vec{x}) t} \ket{0}$ and that ${\bf D}=[0]$ is observed,
the likelihood function for this experiment is given by
\begin{align}
    \label{eq:rabi-like}
    P({\bf D}|\vec{x}; t) & = |\bra{D} e^{-i \omega \sigma_x t/2} \ket{0}|^2
                            = \begin{cases}
                                  \cos^2(\omega t/2) & \text{if ${\bf D} = 0$} \\
                                  \sin^2(\omega t/2) & \text{if ${\bf D} = 1$}
                              \end{cases}.
\end{align}
The likelihood function is then used to update the user's prior beliefs, conditioned on the observed data, using Bayes' theorem:
\begin{align}
    P(\vec{x} | {\bf D}; t) = \frac{
        P({\bf D} | \vec{x}; t) P(\vec{x})
    }{
        \int P({\bf D} | \vec{x}; t) P(\vec{x}) \dd x
    }.
\end{align}
The probability density $P(\vec{x}|{\bf D}; t)$ is known as the posterior distribution, and can be interpreted as the probability distribution that an experimenter should hold after being confronted with data ${\bf D}$ given their preconceptions, which are represented through the prior $P(\vec{x})$.

Except in a few special cases, such as when conjugate priors are used, exact
Bayesian inference is intractable. This is often addressed by using Sequential
Monte Carlo (SMC)~\cite{doucet_tutorial_2011} approximations, also known as particle filters. SMC works by approximating the probability
density using a convex combination of Dirac-delta functions.  In particular,
the probability density at each step is approximated by
\begin{equation}
    P(\vec{x}) \approx \sum_{j=1}^{N_{\rm part}} w_j \delta(\vec{x}-\vec{x}_j).
\end{equation}
This notably reduces the integral in Bayes' theorem to a discrete sum over
$N_{\rm part}$ discrete hypotheses, each of which is called a particle. The
$w_j$ are positive real numbers called weights, and satisfy $\sum_j w_j =1$.
Thus the SMC approximation can be interpreted as a discrete probability
distribution.

A drawback of SMC is that as the algorithm proceeds, the vast majority of the
weights tend to zero.  This is because, with high probability, none of the
initial particles correspond to the true hypothesis. Thus, in the case of Bayesian
inference, the particle
filter approximation above will require a number of particles that is exponential in the length of $\vec{x}$ to estimate
the true hypothesis within fixed error.

The exponential scaling of na\"ive particle filter approximations can be addressed
by moving the particles and changing their weights during the inference
process.  This allows us in many cases to achieve an arbitrarily accurate approximation of the true model with a small (in some cases constant) number of particles. Conventionally, this
is done by \emph{resampling} particles to draw a new set of particles
representing the same posterior distribution.
Resampling algorithms exploit a duality in the SMC
approximation: any probability density can be described either by choosing the
weights of particles appropriately or by moving them such that their
concentrations represent the probability density.  Resampling takes the latter
tack.  It resamples the particles from a new distribution that maintains
certain structural features of the posterior and assigns all the new particles
to have the same weight.  This in effect re-concentrates the particles in
regions where higher probability regions while removing them from places where
the probability is low.

One such resampler that has gained great popularity for physics applications is the Liu--West resampler~\cite{liu_combined_2001}.  The Liu--West resampler draws $N_{\rm part}$ particles from the probability distribution $P(\vec{x})$.  The mean $\mu$ and covariance matrix $\Sigma$ are then computed for the SMC approximation.  The resampler then, for parameter $a\in [0,1]$, shifts the particle slightly towards the mean of the SMC distribution, letting $\mu_j \gets a\vec{x}_j + (1-a)\mu$.  Finally it draws a new particle from the distribution $\mathcal{N}(\mu_j, [1-a^2]\Sigma)$ and assigns a weight of $1/N_{\rm part}$ to the particle.

In cases where $a = 1$, Liu--West resampling is equivalent to the bootstrap filter
\cite{doucet_introduction_2001}, in which each new particle is drawn from
the original SMC distribution with replacement.
If $a = 0$, the resampler simply draws particles from a Gaussian that matches the posterior distributions mean and covariance,
as is useful for rejection filtering \cite{wiebe2015bayesian}.
Typically $a = 0.98$ works well in practice~\cite{liu_combined_2001}; although the optimal value can depend on the likelihood function.  Furthermore, for any $a\in [0,1]$, the distribution of accepted samples has the same mean and covariance as the initial SMC approximation.  Thus the Liu--West resampler is specifically designed to preserve the first two moments, keeping much of the structure-preserving features of the bootstrap filter, while allowing a model to be estimated with exponentially fewer particles
than the bootstrap filter requires.

\begin{figure}
    \begin{center}
        \includegraphics[width=0.8\textwidth]{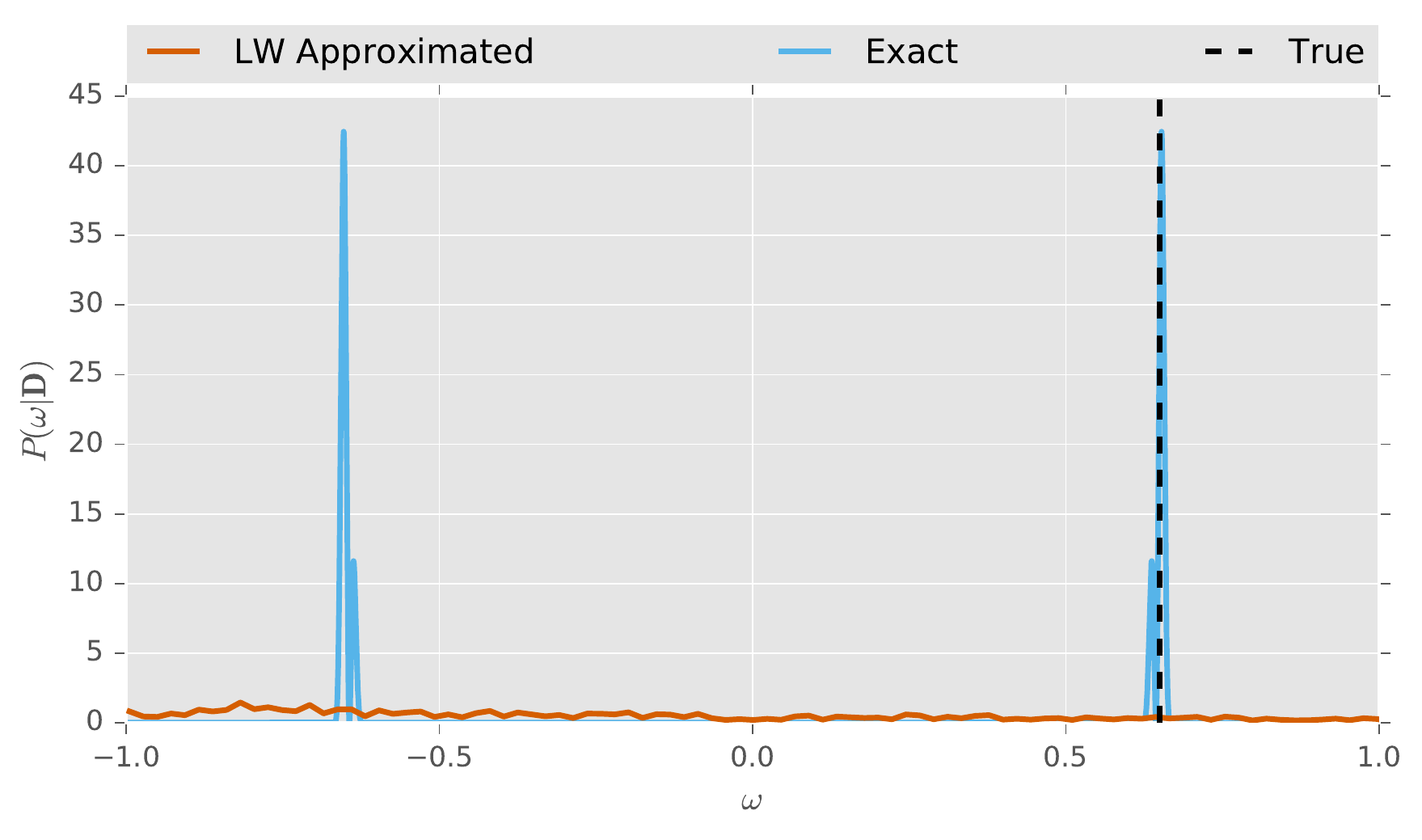}
    \end{center}
    \caption{
        \label{fig:lw-resampling-failure}
        Failure of Liu-West resampling to capture multi-modal posterior
        for the Rabi likelihood \autoref{eq:rabi-like}. The LW-approximated
        and exact posteriors are shown for 40 single-shot measurements at times
        $t_k = (9 / 8)^k$ following the suggestion of \citet{ferrie_how_2013},
        with a uniform prior on $\omega \in [-1, 1]$.
    }
\end{figure}
\begin{figure}
    \begin{center}
        \includegraphics[width=0.8\textwidth]{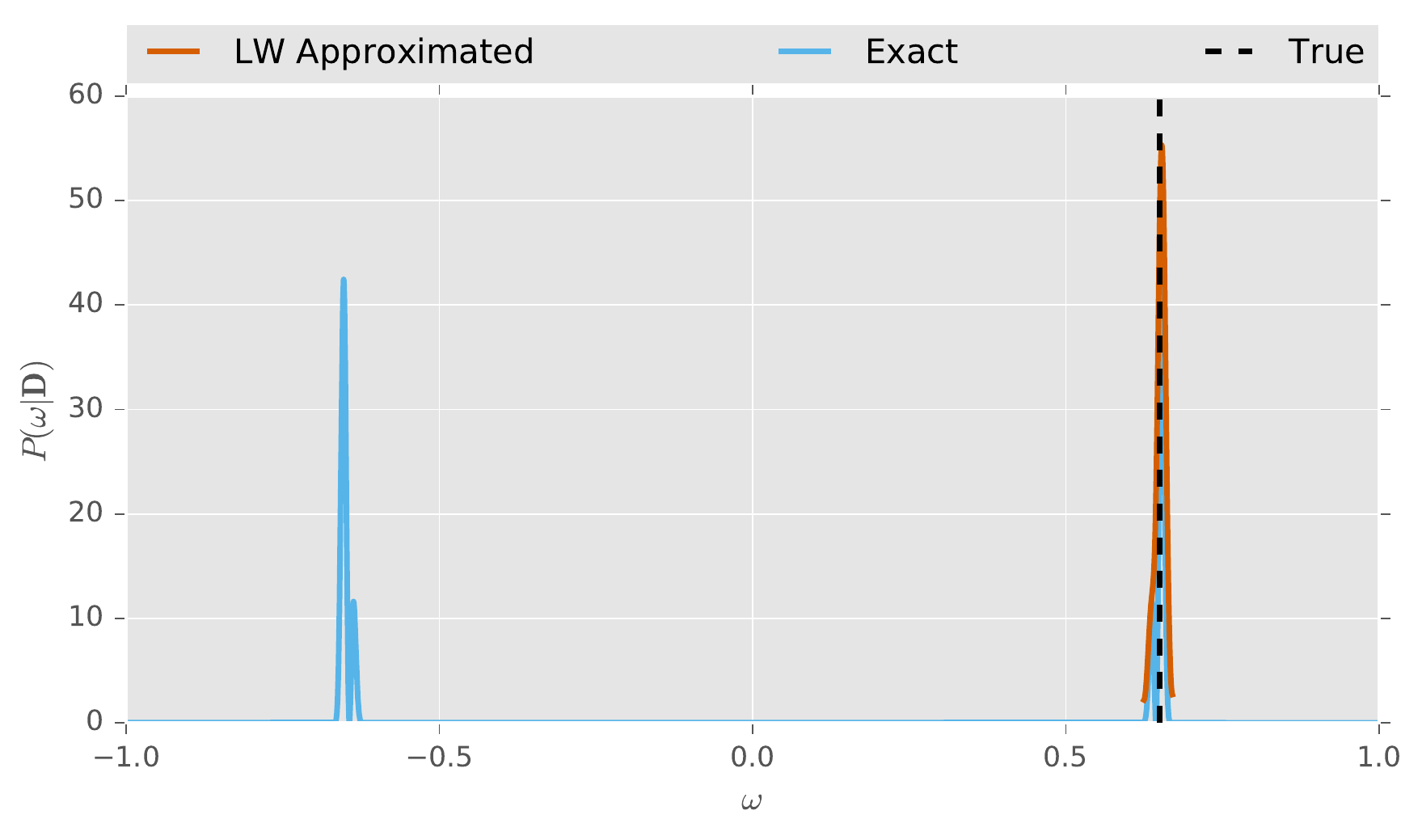}
    \end{center}
    \caption{
        \label{fig:lw-resampling-success}
        Success of Liu-West resampling for capturing the unimodal posterior
        for the Rabi likelihood \autoref{eq:rabi-like}. The LW-approximated
        and exact posteriors are shown for 40 single-shot measurements at times
        $t_k = (9 / 8)^k$, with a uniform prior on $\omega \in [0, 1]$.
    }
\end{figure}

The Liu--West resampling algorithm works in a wide range of applications~\cite{buckland2004state,golightly2011bayesian,granade_robust_2012,wiebe2014hamiltonian}, providing a practical means of implementing Bayesian inference in
experimental practice. There is a major drawback to this approach though, in
that the benefits of the Liu--West algorithm do not extend to multi-modal
distributions. By choosing to perturb the resampled particles towards the mean
of the SMC distribution, we implicitly bias the distribution towards a
unimodal posterior.  In cases where the distribution is multi-modal this
strategy can fail horribly because very little probability density is actually
located near the mean.  We see this in~\autoref{fig:lw-resampling-failure},
where we consider learning the Rabi frequency given by~\autoref{eq:rabi-like},
using Liu-West resampling for $\omega\in [-1,1]$.  In this case, the
likelihood function assigns equal values for both positive and negative
frequencies.  We refer to likelihood functions that have such symmetries as
\emph{degenerate}.  This means that regardless of the true frequency, exact
Bayesian inference on a uniform prior will always yield a bimodal distribution
with zero mean.  Since the Liu--West resampler always moves particles towards
the mean, we expect Liu--West resampling to fail \cite{granade_characterization_2015}
and indeed notice such a failure in the numerical experiments
given in~\autoref{fig:lw-resampling-failure}.

In some ways this is a trivial problem, since the experiment cannot possibly
distinguish between positive and negative frequency.  Thus, if the user is
aware of this degeneracy, they can choose $\omega\in [0,1]$ and learn the sign
in subsequent experiments.  We see that this approach is successful in~\autoref{fig:lw-resampling-success}.  However, not all degeneracies are so obvious or so
easily countered. For example, the degeneracies that appear in randomized gap
estimation (RGE)~\cite{zintchenko_randomized_2016} are significantly harder to
incorporate than in the Rabi example. Rather than placing the onus of data
processing on the user, it is highly desirable to have a method for
automatically addressing such problems as they appear.

Similar challenges emerge when learning with nearly degenerate likelihood
functions. Current practices for mitigating this  include
using significantly more particles and less frequent resampling
steps, avoiding transient failures of the Liu--West
algorithm~\cite{struchalin2016experimental}.
In such cases, Liu--West resampling will ultimately be successful once enough data has been accumulated to break the degeneracy.
However, until the degeneracy can be resolved the algorithm needs to keep track of all the potential hypotheses that could explain the distribution,
mandating a much larger number of particles.
Liu--West resampling will often fail long before it is able to break the degeneracy.  Such problems are often addressed by using an approach known as \emph{annealing} (which is a distinct concept from simulated annealing) with multiple restarts~\cite{del2006sequential}, however such approaches do not give a good estimate of the posterior distribution and require substantial fine tuning.  Instead, it would be very useful to have a resampler that infers and preserves the actual structure of the posterior distribution without requiring domain knowledge.  In the next Section,
we detail our algorithm for finding appropriate mixture models and applying
resampling on the resultant structure in a way that preserves even highly multi-modal
posteriors.

\section{Structured Filtering}
\label{sec:structured}

How does one discover the structures that need to be preserved in the posterior distribution to allow resampling to succeed for degenerate distributions?  One simple approach is to
cluster the posterior using an algorithm like $k$-means clustering~\cite{mackay2003information}; however, unless we have domain knowledge it may not be clear a priori how many clusters to use.  Here we take a somewhat bold step and use an AI-based approach that allows a computer to entertain multiple potential clusterings and reason about which clustering is best.  By allowing the algorithm to decide upon the structures that are most relevant and
are most parsimonious with future data, the problem of dealing with spurious near-degeneracies can be avoided.

We discuss the components of this AI-based solution below.  First, we discuss clustering.  Second, we discuss our method for selecting a model for the posterior.  Third we discuss how clustering and selection can be combined together in a single graphical framework.  Finally we discuss the details of how our structured filtering algorithm combines these features to learn a model for the posterior distribution.

\subsection{Weighted and Unweighted Clustering}
\label{sec:clustering}

\begin{figure}
    \begin{center}
        \includegraphics[width=0.4\textwidth]{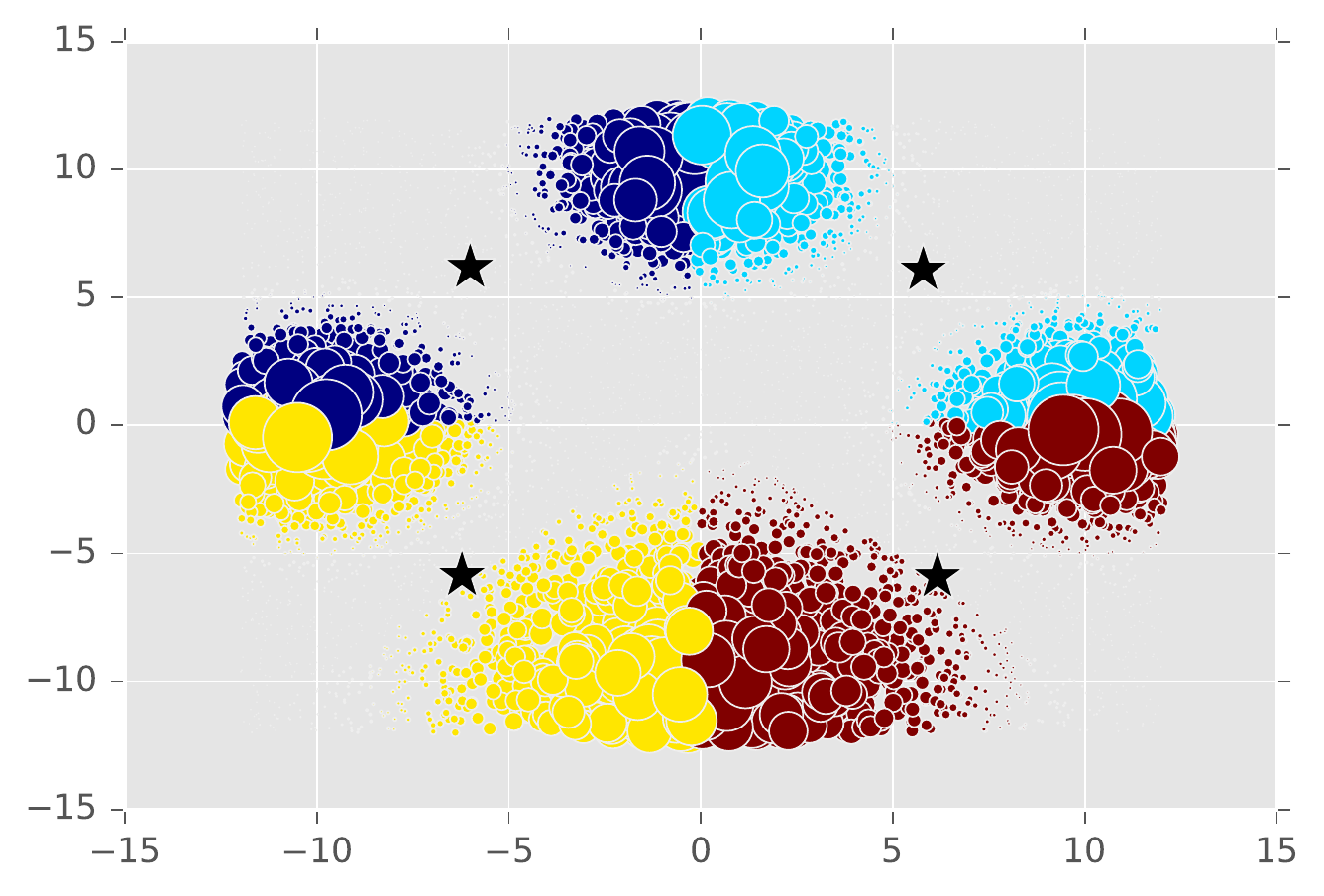}
        \includegraphics[width=0.4\textwidth]{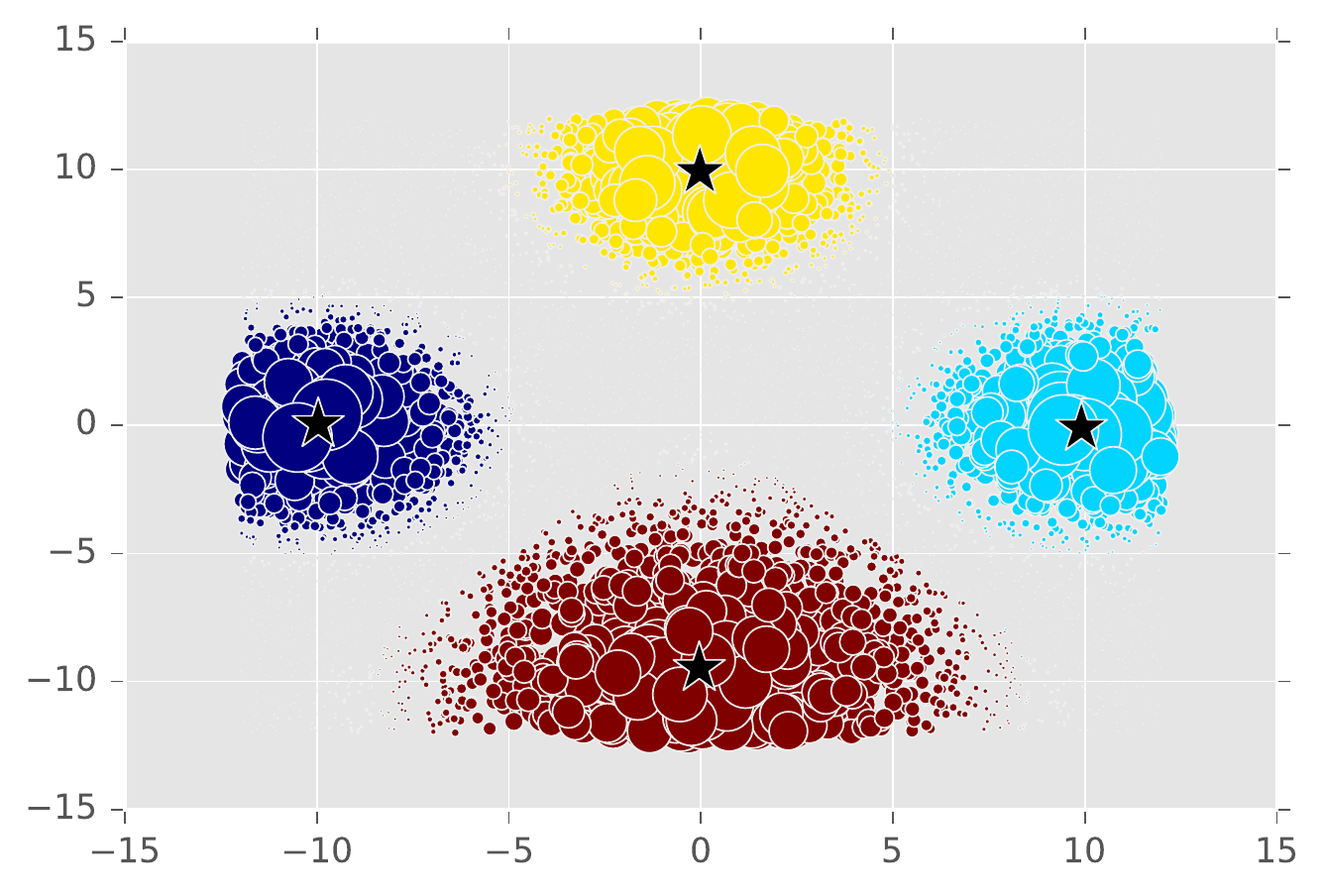}
    \end{center}
    \caption{
        \label{fig:clustering-unweighted-v-weighted}
        Comparison of unweighted (left) and weighted (right) $k$-means
        clustering on 5,000 weighted particles. The positions of
        each particle were drawn uniformly at random, while the
        weights were chosen to represent a mixture of four Gaussian distributions.
        The magnitude of each weight is visualized by the size of the points in the figure,
        with very low-weight particles being ommited for visual clarity.
        The centroids found by each algorithm are indicated by stars.
        The unweighted clustering was performed using SciKit-Learn
        \cite{pedregosa_scikit-learn:_2011},
        while the weighted clustering was performed using \autoref{alg:weighted-kmeans}.
    }
\end{figure}

Clustering seeks to solve a problem that is often second nature to humans: clustering data points into groups of related examples.  While humans excel at this in low dimensions, getting computers to effectively cluster data with the same robustness that humans exhibit can be comparably challenging.  Perhaps the most popular algorithm for clustering data is the $k$-means clustering algorithm.  The algorithm seeks to divide a set of $N$ points into $k$ clusters that minimizes the intra-cluster variances of the clusters.  Specifically we define $S_p$ to be the $p^{\rm th}$ cluster, and then seek cluster centroids $\{\vec{y}_p:p=1,\ldots,k\}$ to satisfy
\begin{equation}
    \label{eq:kmeans}
    \{\vec{y}\}= {\rm argmin}\left(\min_{S_{1},\ldots,S_{k}}\sum_{p=1}^k \sum_{\vec{x}_j \in S_p}\|\vec{x}_j - \vec{y}_p\|_2^2\right).
\end{equation}

An exact solution to~\autoref{eq:kmeans} is unlikely to be found in general.  In fact, finding an optimal cluster assignment is known to be \NP-complete, which means that the existence of any efficient algorithm for finding the optimal clusters would imply $\P=\NP$.  Since it is widely conjectured that $\P \ne \NP$, it is fair to assume that $k$--means clustering is hard in general.  Despite the apparent difficulty of such clustering problems, it can be shown that small perturbations about the hard instances can render them efficiently solvable~\cite{mahajan2009planar}.  Thus the average complexity of clustering is polynomial, which is why at least approximate clustering is not a computationally challenging task.

The $k$-means clustering algorithm is simple.  Given $k$ clusters of vectors compute the centroids of each cluster and assign $\{\vec{y}\}$ to these values.  For each vector (data point) in the set find the cluster centroid that is closest to the example with respect to the Euclidean norm and take each $S_p$ to be the corresponding cluster.  This procedure is then repeated until convergence is reached.

The standard $k$-means clustering algorithm cannot be directly applied to weighted data, such as particles in a SMC approximation, as seen in~\fig{clustering-unweighted-v-weighted}.  The reason for this is that the objective function in~\autoref{eq:kmeans} applies a penalty solely on the distance between the vectors and the centroid.  This is to say that it considers all vectors in the data set equally important, irrespective of their weights.  

Fortunately we can address this by using a \emph{weighted $k$--means} algorithm.  This algorithm instead seeks to minimize
\begin{equation}
\label{eq:weightedkmeans}
    \{\vec{y}\}= {\rm argmin}\left(\min_{S_{1},\ldots,S_{k}}\sum_{p=1}^k \sum_{\vec{x}_j \in S_p} w_j\|\vec{x}_j - \vec{y}_p\|_2^2\right).
\end{equation}
The weighted $k$--means algorithm proceeds exactly as the unweighted version except the cluster centroids are computed as the expectation values of the vectors in each cluster (after renormalizing the weights into a probability distribution) rather than simply by summing the results.  We note in~\fig{clustering-unweighted-v-weighted} that this modification allows multi-modal weighted data to be appropriately divided into clusters.  A formal algorithm for this procedure is given as~\alg{weighted-kmeans} in \apx{pseudocode}.

We utilize the weighted $k$-means algorithm with the
$k$-means++ heuristic for initial centroid selection \cite{arthur_k-means++:_2007} to divide our cluster into a set of different clusters.  In practice,
practitioners will often choose which value of $k$ to use in modeling distributions
by plotting the objective function \autoref{eq:kmeans} and choosing
an inflection point, as such inflection points are suggestive of
overfitting.
By contrast, model selection provides a more formal approach to reasoning
about overfitting \cite{akaike_likelihood_1980}.
In particular, below we discuss the Bayes factor \cite{edwards_bayesian_1963},
which allows us to algorithmically decide on the correct cluster number $k$
for application to posterior distributions.

\subsection{Model selection for the posterior}

Ultimately, the task of selecting the number of clusters to use to represent a SMC approximation to a posterior is one of model selection.  For example, we could have one model that uses $k=2$ another that uses $k=4$ and we wish to know which model (i.e. clustering) does a better job of representing the data.  In this case there's a straight forward approach: consider both clusterings for the posterior and only make a decision between the two when sufficient evidence mounts for the superiority of one of the two models.

We use Bayes factors to compare the validity of two models.  The Bayes factor can be thought of as a generalization of the likelihood ratio test, and is defined for models $M_1$ and $M_2$ and data record $D$ as
\begin{align}
    K = \frac{P(M_1 | D)}{P(M_2 | D)}
      = \frac{P(D | M_1) P(M_1)}{P(D | M_2) P(M_2)}
      = \frac{P(M_1) \int P(\vec{x_1}|M_1)P(D|\vec{x}_1; M_1) \mathrm{d} \vec{x}_1}{P(M_2) \int P(\vec{x_2}|M_2)P(D|\vec{x}_2; M_2) \mathrm{d} \vec{x}_2}.\label{eq:Kbayes}
\end{align}
Here $K>1$ implies model $1$ is superior to model $2$ and vice versa.
It reduces to the likelihood ratio test when the prior over models is uniform
($P(M_1) = P(M_2)$), and when priors within each model are chosen such that
\begin{align*}
    P(\vec{x}_1) = \delta\left(\vec{x}_1 - {\rm argmax}(P(D|\vec{x},M_1))\right),
\end{align*}
and similarly for the second model.  However, Bayes factors have a very nice feature absent from the likelihood ratio test.  The values of the integrals in~\autoref{eq:Kbayes} depend on the volumes of the parameter spaces of the models.  This penalizes models with more parameters, and gives us a simple alternative to model selection that does not involve maximization as in the Bayesian information criteria~\cite{schwarz1978estimating}.  This means that Bayes factors can easily be computed using SMC approximations from the particle weights and the likelihood function~\cite{ferrie2014quantum}.

In practice, we also use Bayes factors to choose one option among several possible competing options.  In such cases, we compare the values of $K$ for all possible pairs of models that we want to compare and then select a model once $\max_k(\min_{j} K(M_j,M_k)) \ge K_{\rm champ}$ where $K_{\rm champ}$ is a user specified threshold.  Typically a value of $K_{{\rm champ}}\ge 100$ is considered strong evidence in favor of one of the models.  In practice, we usually take $K_{{\rm champ}}\ge 2000$ to be our threshold.  We do this because excluding the correct number of clusters can often have a worse impact on the algorithm performance than entertaining a hypothesis than is likely false and because we perform many such tests in structured filtering.


\subsection{Graphical Models for Posteriors}
\label{sec:graphical-models}

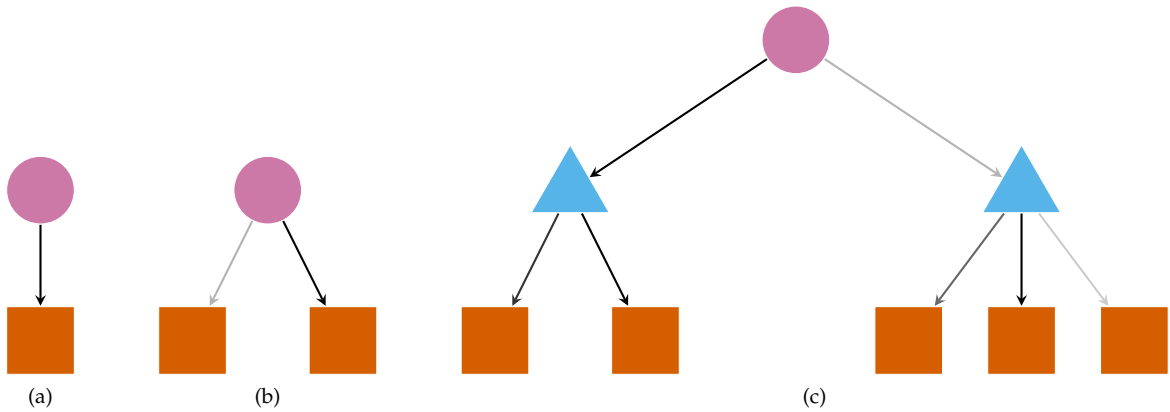
\begin{figure}
    \begin{center}
        \subfloat[]{
            \begin{tikzpicture}[]
                \node[model selection node] (root) at (0, 2) {};
                \node[particle filter node] (a) at (0, 0) {};
                \draw[tree edge] (root) to (a);
            \end{tikzpicture}
        }
        \hskip3em
        \subfloat[]{
            \begin{tikzpicture}
                \node[model selection node] (root) at (0, 2) {};
                \node[particle filter node] (a) at (-1, 0) {};
                \node[particle filter node] (b) at (+1, 0) {};

                \draw[tree edge, black!30] (root) to (a);
                \draw[tree edge] (root) to (b);
            \end{tikzpicture}
        }
        \hskip3em
        \subfloat[]{
            \begin{tikzpicture}
                \node[model selection node] (root) at (0, 4) {};

                \node[mixture model node] (mix2) at (-3, 2) {};
                \node[particle filter node] (a2) at (-4, 0) {};
                \node[particle filter node] (b2) at (-2, 0) {};

                \node[mixture model node] (mix3) at (3, 2) {};
                \node[particle filter node] (a3) at (4.5, 0) {};
                \node[particle filter node] (b3) at (3, 0) {};
                \node[particle filter node] (c3) at (1.5, 0) {};

                \draw[tree edge] (root) to (mix2);
                \draw[tree edge, black!80] (mix2) to (a2);
                \draw[tree edge] (mix2) to (b2);

                \draw[tree edge, black!30] (root) to (mix3);
                \draw[tree edge, black!20] (mix3) to (a3);
                \draw[tree edge, black!100] (mix3) to (b3);
                \draw[tree edge, black!60] (mix3) to (c3);
            \end{tikzpicture}
        }

        \caption{
            \label{fig:graphical-model-examples}
            Examples of graphical models for (a) conventional particle
            filtering, (b) model selection between particle filters using
            distinct likelihood functions, and (c) model selection between
            two different clusterings of the posterior into mixture models.
            In each example, particle filtering is indicated by square
            nodes, model selection is indicated by circle nodes, and mixture
            models are indicated by triangle nodes. The shades of each edge
            correspond to their \emph{weights} with darker edges corresponding to higher weight.
        }
    \end{center}
\end{figure}

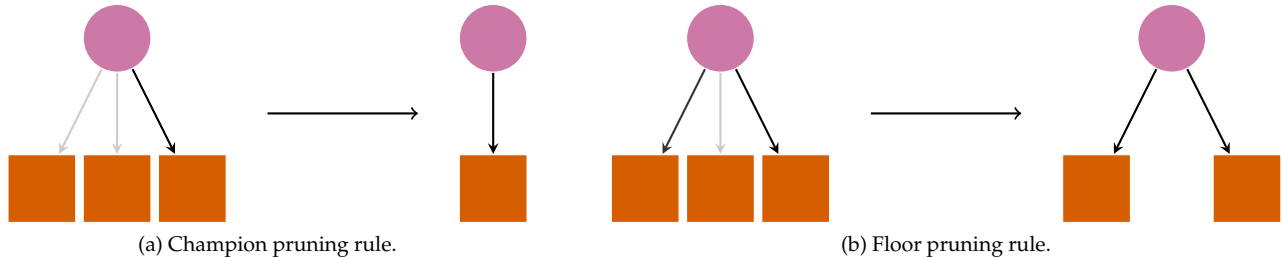
\begin{figure}
    \begin{center}
        \subfloat[Champion pruning rule.]{
            \label{subfig:champion-pruning}
            \begin{tikzpicture}
                    \node[model selection node] (left root) at (-5, 3) {};
                    \node[particle filter node] (left leaf 1) at (-4, 1) {};
                    \node[particle filter node] (left leaf 2) at (-5, 1) {};
                    \node[particle filter node] (left leaf 3) at (-6, 1) {};
                    
                    \draw[tree edge]           (left root) to (left leaf 1);
                    \draw[tree edge, black!20] (left root) to (left leaf 2);
                    \draw[tree edge, black!20] (left root) to (left leaf 3);

                \draw[->, thick] (-3, 2) to (-1, 2);

                    \node[model selection node] (right root) at (0, 3) {};
                    \node[particle filter node] (right leaf 1) at (0, 1) {};
                    
                    \draw[tree edge]           (right root) to (right leaf 1);
            \end{tikzpicture}
        }
        \hskip3em
        \subfloat[Floor pruning rule.]{
            \label{subfig:floor-pruning}
            \begin{tikzpicture}
                    \node[model selection node] (left root) at (-5, 3) {};
                    \node[particle filter node] (left leaf 1) at (-4, 1) {};
                    \node[particle filter node] (left leaf 2) at (-5, 1) {};
                    \node[particle filter node] (left leaf 3) at (-6, 1) {};
                    
                    \draw[tree edge]           (left root) to (left leaf 1);
                    \draw[tree edge, black!20] (left root) to (left leaf 2);
                    \draw[tree edge, black!80] (left root) to (left leaf 3);

                \draw[->, thick] (-3, 2) to (-1, 2);

                    \node[model selection node] (right root) at (1, 3) {};
                    \node[particle filter node] (right leaf 1) at (0, 1) {};
                    \node[particle filter node] (right leaf 2) at (2, 1) {};
                    
                    \draw[tree edge]           (right root) to (right leaf 1);
                    \draw[tree edge]           (right root) to (right leaf 2);
            \end{tikzpicture}
        }

        \caption{
            \label{fig:pruning-rules}
            Examples of the champion and floor pruning rules applied to a
            model selection node with three initial child particle filter nodes.  All such replacement rules work identically if the filter node is replaced with a subtree.
        }
    \end{center}
\end{figure}

In order to convert the problem of automatically choosing the optimal number of clusters in the data into one that a computer can easily solve, we introduce a graphical model for describing the reasoning process that is employed to decide between different clusterings of the posterior distribution.  The graphs we consider are directed rooted trees with edges pointed away from the root node, but more general directed acyclic graphs could also be considered. We provide examples of our graphical notation in
\autoref{fig:graphical-model-examples}.

The vertices in these graphs are given one of three distinct labels, in addition to their index.   The first such label denotes that the vertex is a filter node, which serves as a container for a set of SMC particles that we assume are approximately unimodal.  These nodes are denoted by squares.  It is important to note that the filter nodes do not necessarily need to have the same likelihood functions,
allowing for the inclusion of more general model selection problems.  Furthermore, we do not need to use Liu--West resampling inside each of the filter nodes.  Other filters such as the bootstrap filter, assumed density filtering~\cite{minka2001expectation} or rejection filtering~\cite{wiebe2015bayesian} can be used in its place.

The second type is a mixture node.  The mixture node defines a distribution that is a weighted mixture of the subtrees that descend from it, but contains no particles itself.  For example, a $2$ cluster approximation to the posterior would be described by a mixture node with two filter nodes as leaves.  We denote mixture nodes as triangles.  The ability to mix multiple nodes has several interesting properties.  First, in principle we can emulate a filter node consisting of many particles with a mixture of single-particle filter nodes.  Second, we can have mixtures of mixtures.  This allows us to generate very rich clusterings even if the maximum degree for the graph is $3$ (that is, if
we restrict each node to have at most $2$ children and one parent).

The final label that a vertex can be assigned corresponds to a decision node.  A decision node is a mixture node but is put in place explicitly to test between the hypotheses described by the subtrees that descend from it.  Such nodes are denoted with a circle. By convention, the root node is always a decision node
.
The edges in our graphical model describe the relationships between the different types of nodes.  By definition a filter node has no children, and is hence always a leaf node; however, mixture nodes and decision nodes must have children.  The edges between any two nodes are used to assign weights.  In a mixture node these weights are used to set the weights properly for the particles that reside in the filter nodes within the subtrees that descend from them.  In particular, the actual weights are the products of the weights within the filter nodes and all outgoing edges of mixture nodes that connect it to the root.  In a decision node these weights serve to represent the algorithm's confidence that one of the competing hypotheses, described by the descendant subtrees, is correct. 

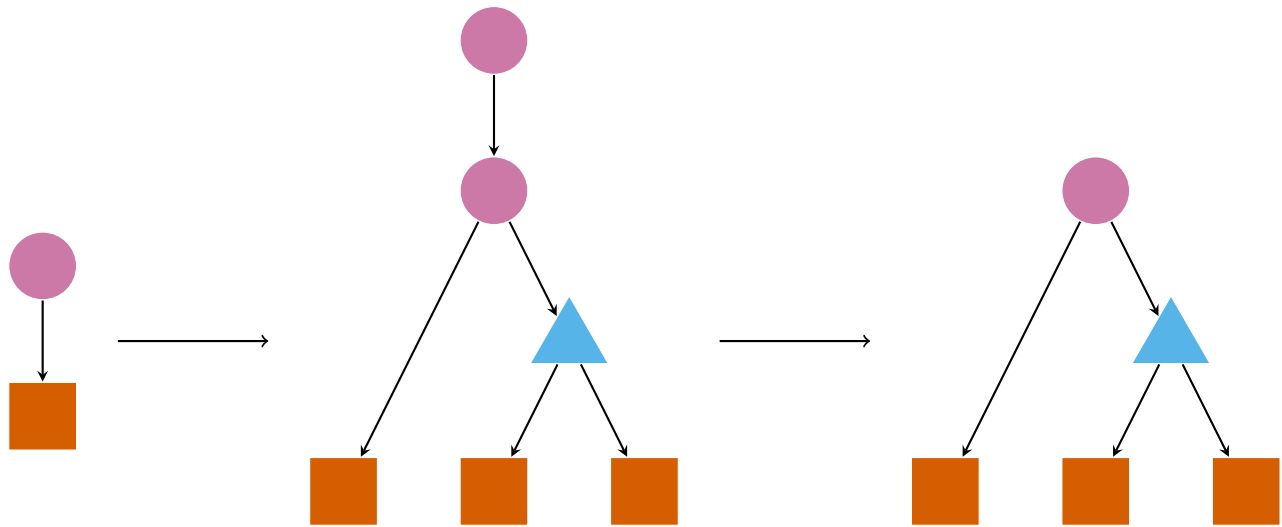
\begin{figure}
    \begin{center}
        \begin{tikzpicture}
                \node[model selection node] (left root) at (-4, 3) {};
                \node[particle filter node] (left leaf) at (-4, 1) {};
                \draw[tree edge] (left root) to (left leaf);

                \draw[->, thick] (-3, 2) to (-1, 2);

                \node[model selection node] (right root) at (2, 6) {};
                \node[model selection node] (right root2) at (2, 4) {};
                \node[particle filter node] (single filter) at (0, 0) {};
                \node[mixture model node] (mix2) at (3, 2) {};
                \node[particle filter node] (mix2 left) at (2, 0) {};
                \node[particle filter node] (mix2 right) at (4, 0) {};

                \draw[tree edge] (right root) to (right root2);
                \draw[tree edge] (right root2) to (single filter);
                \draw[tree edge] (right root2) to (mix2);
                \draw[tree edge] (mix2) to (mix2 left);
                \draw[tree edge] (mix2) to (mix2 right);

                \draw[->, thick] (5, 2) to (7, 2);

                \node[model selection node] (right root2b) at (2+8, 4) {};
                \node[particle filter node] (single filterb) at (0+8, 0) {};
                \node[mixture model node] (mix2b) at (3+8, 2) {};
                \node[particle filter node] (mix2 leftb) at (2+8, 0) {};
                \node[particle filter node] (mix2 rightb) at (4+8, 0) {};

                \draw[tree edge] (right root2b) to (single filterb);
                \draw[tree edge] (right root2b) to (mix2b);
                \draw[tree edge] (mix2b) to (mix2 leftb);
                \draw[tree edge] (mix2b) to (mix2 rightb);
        \end{tikzpicture}

        \caption{
            \label{fig:structured-resampling}
            An example of a \emph{splitting} move, in which
            a single particle filter node is replaced by a model selection
            over different clusterings, each of which is represented by
            an appropriate mixture model. In this example, the structured
            filter considers $n_{\textrm{clusters}} \in \{1, 2\}$; since the
            $n_{\textrm{clusters}} = 1$ mixture model node would be redundant,
            it is immediately eliminated, promoting its only particle filter
            child to be a child of the root model selection node. Similarly,
            the new decision node over the number of clusterings is redundant
            with the root decision node, and is also eliminated immediately.
        }
    \end{center}
\end{figure}

The inference process will often uncover structure in the posterior that was not apparent in earlier steps of the learning process.  This necessitates adapting our graphs in order to match the changing structure.  We do this using three simple rules, which we demonstrate in~\autoref{fig:pruning-rules} and~\autoref{fig:structured-resampling}.  The champion rule and floor pruning rules are designed to simplify the graph when certain structures are not needed to represent the data.  The champion rule states that when the weight of one edge overwhelms the sum total of the other weights sufficiently then all other hypothetical subtrees are disregarded except for the ``champion.''  The floor prune rule immediately eliminates a subtree if the edge weight is smaller than a threshold.  This rule is useful because it allows the algorithm to free up memory and processing to address more fruitful clusterings when one has been effectively excluded by the data.

The splitting move, on the other hand, increases the complexity of the graph.
It takes a particle filter node and replaces that node by a subtree, as
illustrated in \autoref{fig:structured-resampling}.  Specifically, we replace
the node with a decision node that has the original filter node as a child
and also a mixture node that has at least two filter nodes as children.
Although splitting moves could be performed at any time, we 
only deploy them during a resampling step to optimize the performance of the
algorithm.  These moves are implemented in our algorithm by taking the
particle filter in the original node and using weighted $k$--means clustering,
using weighted $k$--means++ to divide it into $2$ or more clusters.  This
guarantees that the new mixture of particle filters is equivalent to the
initial particle filter before resampling is applied. After splitting, each
particle filter leaf node is then resampled independently. Thus, in the
branch of the decision node corresponding to the correct number
of clusters,
the resampling takes place only locally within each cluster and preserves
the multi-modal structure of the posterior.

A complication comes in with the number of particles assigned to each cluster.  By default we divide the particles into two sub-clusters,
assigning each original particle to one of the new clusters using the
labels returned by the $k$-means algorithm. Thus, we retain the total number of particles through a splitting move.  However,
this presents the danger that a cluster with a small number of particles will become numerically unstable, even if the weight
assigned to that cluster is large. To remedy this,
we also allow the number of particles used to vary dynamically by setting a minimum number of particles in each cluster.
For example, imagine we wish to split a cluster with $1800$ particles into $3$ clusters and we set the minimum number of particles per cluster to be $1000$.  
Then, at least one cluster would ordinarily be assigned less than the minimum number of particles
that we have decided upon. Our algorithm will draw additional particles with correspondingly smaller
weights for such clusters, preserving numerical stability while introducing no new approximations.
This results in $3$ clusters with at least $1000$ particles each.
Since we choose to apply restructuring moves (such as splitting) only when a 
resample would be triggered by the Liu--West resampler, generating additional particles is easy to do by following the same perscription used
in the resampler.

Applied directly, the splitting move would generate exponentially-large trees
for even simple models. We limit this by imposing a maximum depth $d_{\max}$
at which the splitting move may be applied. If a particle filter node with
depth $d \ge d_{\max}$ must be resampled, then we apply traditional Liu--West
resampling at that node instead. In this way, we can control the maximum size
of the structure that our algorithm is allowed to explore.

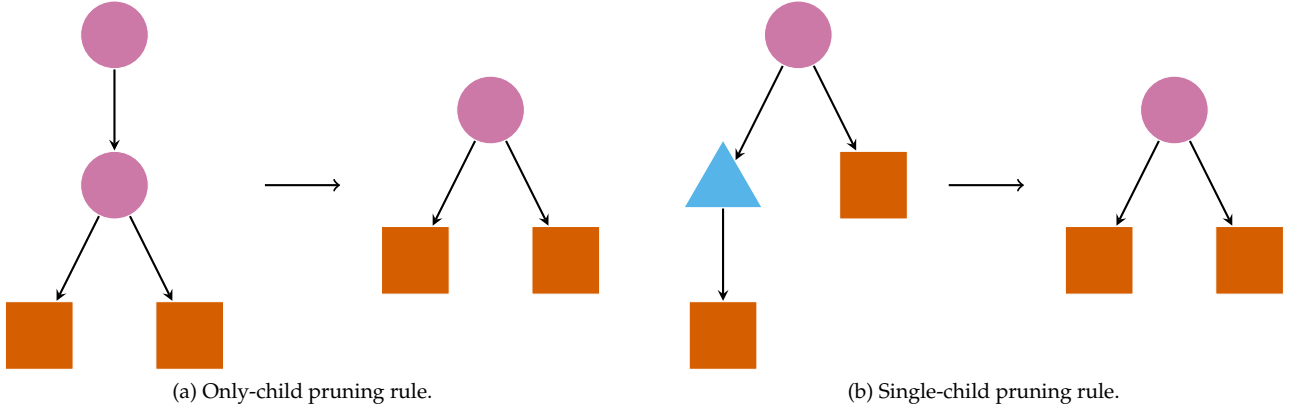
\begin{figure}
    \begin{center}
        \subfloat[Only-child pruning rule.]{
            \label{subfig:only-child}
            \begin{tikzpicture}
                    \node[model selection node] (left root) at (-4, 3) {};
                    \node[model selection node] (left only child) at (-4, 1) {};
                    \node[particle filter node] (left leaf 1) at (-3, -1) {};
                    \node[particle filter node] (left leaf 2) at (-5, -1) {};
                    \draw[tree edge] (left root) to (left only child);
                    \draw[tree edge] (left only child) to (left leaf 1);
                    \draw[tree edge] (left only child) to (left leaf 2);

                    \draw[->, thick] (-2, 1) to (-1, 1);

                    \node[model selection node] (right root) at (1, 2) {};
                    \node[particle filter node] (right leaf 1) at (0, 0) {};
                    \node[particle filter node] (right leaf 2) at (2, 0) {};
                    \draw[tree edge] (right root) to (right leaf 1);
                    \draw[tree edge] (right root) to (right leaf 2);
            \end{tikzpicture}
        }\hskip3em\subfloat[Single-child pruning rule.]{
            \label{subfig:single-child}
            \begin{tikzpicture}
                    \node[model selection node] (left root) at (-4, 3) {};
                    \node[mixture model node] (left single child) at (-5, 1) {};
                    \node[particle filter node] (left leaf) at (-5, -1) {};
                    \node[particle filter node] (left other leaf) at (-3, 1) {};
                    \draw[tree edge] (left root) to (left single child);
                    \draw[tree edge] (left root) to (left other leaf);
                    \draw[tree edge] (left single child) to (left leaf);

                    \draw[->, thick] (-2, 1) to (-1, 1);

                    \node[model selection node] (right root) at (1, 2) {};
                    \node[particle filter node] (right leaf 1) at (0, 0) {};
                    \node[particle filter node] (right leaf 2) at (2, 0) {};
                    \draw[tree edge] (right root) to (right leaf 1);
                    \draw[tree edge] (right root) to (right leaf 2);
            \end{tikzpicture}            
        }

        \caption{
            \label{fig:only-single-child-moves}
            Examples of the only-child and single-child pruning moves, used to
            eliminate redundant intermediate nodes and reduce the depth of the
            structured filtering tree. Each of the pruning
            rules shown above exactly preserve the structure with a simpler tree.
        }
    \end{center}
\end{figure}

Though these three moves are sufficient to correctly implement structured
filtering, we also consider two other pruning moves which reduce graph
complexity without additional approximation. These moves  allow the
algorithm to reduce the depth of the tree dynamically as the floor and
champion rules discussed above eliminate uninformative branches of the tree.
In particular, the only-child and single-child pruning rules shown in
\autoref{fig:only-single-child-moves} replace the current tree with a simpler
tree describing the same structure by eliminating redundant intermediate nodes.
Under the only-child pruning rule, a node is eliminated if it is the only
child of its parent and has one or more children, such that those children
can be attached directly to its parent. Similarly, the single-child rule
removes any node with exactly one child, and places that child directly
onto its grandparent. Trees matching the preconditions for only-child
and single-child pruning are generated by applying the champion and
floor pruning rules, as each of those eliminate leaf and intermediate
notes that do not contribute substantially to the final estimate.
By removing these intermediate nodes, the depth of nodes relevant to
the final estimate can be decreased, allowing for the splitting move to
be applied again.

Importantly, the floor, champion, and splitting moves are each
\emph{parameterized}, offering quality parameters for the particle filter
approximation represented by the output of these moves. To allow these pruning
and splitting parameters to dynamically depend on the tree structure, we
encode them as properties of each node, collectively called a \emph{context}.
In this way, our algorithm can be customized with various starting trees
representing prior knowledge about initial clustering, model selection
problems of interest, or other structure of interest. Structured filtering recognizes
the following context parameters, each of which may be specified at a given
node, or \emph{inherited} from the a node's parent:

\begin{description}
    \item[Prune (boolean)] If this context parameter is set to false, then
        no pruning is applied at this node; in particular, no floor, champion,
        only-child or single-child rules are applied. For brevity, we omit this
        context parameter in the algorithm below.
    \item[Mixture floor (real)] This context parameter sets the
        minimum edge weight from a mixture node to one of its children
        that will be preserved by floor pruning.
    \item[Decision floor (real)] This context parameter sets the
        minimum edge weight from a decision node to one of its children
        that will be preserved by floor pruning.
    \item[Decision champion $K_{\mathrm{champ}}$ (real)] This context parameter sets the
        ratio by which the edge weight from a decision node to one of its
        children must exceed the sum of all other outgoing edge weights from
        the same decision node before that child will be considered a champion
        node.
    \item[Decision region estimation champions (integer)] This
        context parameter controls the number of children of each decision node
        that will be kept when reporting region estimates, as
        described in \autoref{sec:sf-region-est}.
\end{description}

\subsection{Structured Region Estimation}
\label{sec:sf-region-est}

As described by \citet{ferrie_high_2014}, clustering can also be used to report
region estimates of higher posterior density than unclustered methods. We use
and generalize this observation by exploting the tree structure generated by
splitting and pruning moves to form powerful credible region estimators.
In particular, each particle filter (leaf) node already yields conventional
region estimators such as covariance ellipsoids, convex hulls, and minimum
volume enclosing ellipsoids \cite{granade_robust_2012}, such that we can
complete our region estimation procedure by specifying region estimators
for each decision and mixture node, recursively.

Following this strategy, at each mixture node, our estimation procedure
assigns a region estimate that is the union of the region estimates for each
of its children. At each decision node, our procedure reports the union of the
first $n$ of its children's region estimates, with its children arranged in
descending order by their weights, and with $n$ obtained from the
corresponding context parameter, as described in
\autoref{sec:graphical-models}.
The final region estimate $X_\alpha$ can thus be interpreted as
guaranteeing that the probability the model vector $\vec{x}$ is within $X_\alpha$,
conditioned on the model with the highest posterior probability, is at least
$\alpha$. Importantly, the credibility parameter $\alpha$ is only used at the
leaf nodes as a parameter to the local region estimation procedure.

\subsection{Structured Filtering Algorithm}

Our algorithm depends on two global parameters as well as the context
parameters described above.

\begin{description}
    \item[$d_{\max}$ (integer)] This global parameter sets the maximum depth that a node is allowed to be from the root
in the structure graph.
    \item[$\{n_{\mathrm{cluster}, i}\}$ (set of integers)]
        This global parameter sets degrees of each vertex in the structure graph that will be
        considered by splitting moves. This imposes
        a limit on the maximum number of clusters for the data of $(\max_i n_{\mathrm{cluster}, i})^{d_{\max} - 1}$.
\end{description}

With this description in place, we now present our algorithm in full as \autoref{alg:structured-filtering}.
Important subroutines are listed separately in \apx{pseudocode}.
The splitting move
used in \autoref{alg:structured-filtering} is demonstrated graphically in
\autoref{fig:structured-resampling}.

The numerical examples shown in \autoref{sec:results},
\autoref{sec:phase-est} and \apx{numerics} were obtained using an implementation
in Python 2.7 (Anaconda distribution), with the NumPy
\cite{walt_numpy_2011}, SciPy \cite{jones_scipy:_2001}, and QInfer
\cite{granade_qinfer:_2016} libraries.

\begin{algorithm}[H]
    \caption{\label{alg:structured-filtering}
        Structured filtering algorithm.
    }
    \begin{algorithmic}
        \Require
            Number of particles $\npart$,
            minimum number of particles $\minnpart$,
            max depth $d_{\max}$,
            set of cluster numbers to consider $\{n_{\cluster,i}\}$,
            initial context $\mathbf{c}$,
            local resampler $R$,
            local resample threshold $r \in [0, 1]$,
            number of experiments $\nexp$,
            prior distribution $\pi(\vec{x})$,
            likelihood function $\Pr(d | \vec{x})$.
        
        \Function{StructuredFilter}{
            $\npart$, $d_{\max}$, $\{n_{\cluster,i}\}$,
            $\mathbf{c}$, $R$, $D$, $\pi$, $\Pr(d | \vec{x})$
        }
            \seccomment{Initialization.}
            \State Create a new filter node $\phi$ by drawing
                $\npart$ particles from $\pi$.
            \State Create a new decision node $\rho$ with $\phi$ as
                its only child, and assign $\textbf{c}$ as its context.
            \State Set the weight of the edge $\rho \to \phi$ to one.
            \seccomment{Data processing.}
            \For{$\iexp \in \{1, \dots, \nexp\}$}
                \seccomment{Experiment design and data collection.}
                \For{each filter node $\phi$ descended from $\rho$}
                    \State Let $\bfop(\phi)$
                        be the product of the weights of each edge
                        leading from $\rho$ to $\phi$.
                \EndFor
                \State Let $\phi_{\min} = {\arg\,\min}_\phi \bfop(\phi)$
                    be the filter node with minimal Bayes factor.
                \State Draw unique $\{\vec{x}_1, \vec{x}_2\} \sim \phi$.
                \State Let $t = 1 / \|\vec{x}_1 - \vec{x}_2\|$.
                \State Perform the experiment $\vec{e} = (t, \vec{x}_1)$,
                    collecting the outcome $d$.
                \seccomment{Update via tree traversal.}
                \For{each node $\nu$ in a depth-first traversal from $\rho$}
                    \If{$\nu$ is a filter node}
                        \State Let $\{w_i\}$ and $\{\vec{x}_i\}$ be the
                            weights and particles at $\nu$.
                        \State Update each $w_i$ as
                        $$
                            w_i \mapsto w_i \times \Pr(d | \vec{x}_i; \vec{e}).
                        $$
                        \State Multiply the weight of the edge to
                            $\nu$ from its parent by $\sum_i w_i$.
                        \State Renormalize $w_i \mapsto w_i / \sum_i w_i$.
                    \Else
                        \inlinecomment{Push weights up the tree.}
                        \State Multiply the weight of the edge to
                            $\nu$ from its parent by the sum of the edge
                            weights outgoing from $\nu$.
                        \State Renormalize the weights outgoing from $\nu$
                            to sum to 1.
                    \EndIf
                \EndFor
                \seccomment{Pruning.}
                \For{each node $\nu$ descended from $\rho$}
                    \If{$\nu$ has at least two children}
                    \inlinecomment{See also: \autoref{subfig:champion-pruning}.}
                        \State Let $w$ be the largest weight of an
                            edge outgoing from $\nu$.
                        \If{$w / (1 - w) > $ the current context's champion threshold}
                            \State Let $\chi$ be the child of $\nu$ with
                                edge weight $w$.
                            \State Remove all children of $\nu$ except for that
                                $\chi$.
                            \State Set the weight of $\nu \to \chi$ to 1.

                        \EndIf
                    \EndIf
                    \vskip0.25em
                    \If{$\nu$ is a selection node}
                    \inlinecomment{See also: \autoref{subfig:floor-pruning}.}
                        \For{each child $\chi$ of $\nu$}
                            \If{weight of $\nu \to \chi < $ the current context's floor threshold}
                                \State Remove $\chi$.
                            \EndIf
                        \EndFor
                        \State Renormalize the weights outgoing
                            from $\nu$ to sum to 1.
                    \EndIf
                    \vskip0.25em                    
                    \If{$\nu$ is the only child of its parent}
                    \inlinecomment{See also: \autoref{subfig:only-child}.}
                        \State Let $\alpha$ be the parent of $\nu$.
                        \For{each child $\chi$ of $\nu$}
                            \State Move $\chi$ to be a child of $\alpha$,
                                keeping the weight of the current edge
                                $\nu \to \chi$.
                        \EndFor
                        \State Remove $\nu$ from the children of $\alpha$.
                    \EndIf
                    \inlinecomment{\emph{continued on next page}}
                    \algstore{structured-filtering}
    \end{algorithmic}
\end{algorithm}
\begin{algorithm}[H]
    \begin{algorithmic}
                    \algrestore{structured-filtering}
                    \If{$\nu$ has exactly one child}
                    \inlinecomment{See also: \autoref{subfig:single-child}.}
                        \State Let $\chi$ and $\alpha$ be the child and parent of $\nu$.
                        \State Append $\chi$ as a child of $\alpha$, with weight
                            given by the current edge $\alpha \to \nu$.
                        \State Remove $\nu$ from the children of $\alpha$.
                    \EndIf
                \EndFor
                \seccomment{Structured resampling.}
                \For{each filter node $\phi$ descended from $\rho$}
                    \State Let $\{w_i\}$ and $\{\vec{x}_i\}$
                        be the weights and particles at $\phi$.
                    \State Let $n_\ess(\phi) = 1 / \sum_i w_i^2$ be the effective
                        sample size of $\phi$ and $n(\phi)$ be the number
                        of particles at $\phi$.
                    \If{$n_\ess / n(\phi) < r$}
                        \If{depth of $\phi < d_{\max}$}
                            \linecomment{Perform a splitting move (\autoref{fig:structured-resampling})}.
                            \State Replace $\phi$ in its parent by
                                a new decision node $\delta$.
                            \For{$n_{\cluster} \in \{n_{\cluster,i}\}$}
                            \inlinecomment{
                                Make new mixture nodes to represent
                                each possible number of clusters.
                            }
                                \If{$n_{\cluster} = 1$}
                                    \State Append a copy $\phi'$ of $\phi$ to $\delta$.
                                    \State Locally resample $\phi'$ using $R$.
                                \Else
                                    \State
                                        Let $\{\ell_i\}, \{y_j\} = {}$\Call{WeightedKMeans}{
                                            $\{\vec{x}_i\}$, $\{w_i\}$, $n_{\cluster}$
                                        }.
                                        \inlinecomment{See also \autoref{alg:weighted-kmeans}.}
                                    \State Make a new mixture node $\mu$ and append it
                                        as a child of $\delta$.
                                    \For{$j \in \{1, \dots, n_{\cluster}\}$}
                                    \inlinecomment{
                                        Use local resampler to populate new
                                        filter nodes.
                                    }
                                        \State Make a new filter node $\phi_j$ and
                                            append it as a child of $\mu$.
                                        \State Let $I_j =
                                            \{i \in \{1, \dots, n(\phi)\} | \ell_i = j\}$ be the indices of the $j^{\text{th}}$
                                            cluster.
                                        \State Set the weight of $\mu \to \phi_j$
                                            to $\sum_{i \in I_j} w_i$.
                                        \State Draw $\max(\minnpart, |I_j|)$ particles
                                            $\{\vec{x}'_i\}$ from $R(\{w_i : i\in I_j\}), \{\vec{x}_i : i \in I_j\})$.
                                        \State Set the particles at $\phi_j$ to be
                                            $\{\vec{x}'_i\}$, with uniform weights
                                            $1 / |I_j|$.
                                    \EndFor
                                \EndIf
                            \EndFor
                            \State Set the weights of edges outgoing from
                                $\delta$ to be uniform and summing to 1.
                        \Else
                            \State Locally resample $\phi$ with $R$.
                        \EndIf
                    \EndIf
                \EndFor
                \vskip0.5em
            \EndFor
            \vskip0.5em
        \EndFunction
    \end{algorithmic}
\end{algorithm}

\section{Application to randomized gap estimation}
\label{sec:results}

To demonstrate the effectiveness of our structured filtering algorithm,
we consider the randomized gap estimation (RGE) model of
\citet{zintchenko_randomized_2016}. An RGE experiment consists of choosing
a state $\ket{\psi} = U\ket{0}$ for a Haar-random unitary $U$ (or $U$ sampled from a $2$--design) and
fixed preparation $\ket{0}$, evolving under the unknown
Hamiltonian for a time $t$, then measuring
$\{\ket{\psi}\bra{\psi}, \id - \ket{\psi}\bra{\psi}\}$. Labelling
the measurement outcome $\ket{\psi}\bra{\psi}$ as 0, we obtain the likelihood
function for RGE,
\begin{align}
    \label{eq:rge-like}
    \Pr(0 | H; t) = \sum_{\substack{i > j\\ \{i, j\} \subseteq \{1, \dots, k\}}}
        \cos^2([\lambda_i - \lambda_j] t / 2),
\end{align}
where $\dim H = k + 1$ and where $H$ has eigenvalues
eigenvalues $\{\lambda_0, \dots, \lambda_k\}$.
RGE is significant because it gives a way to estimating the gaps
$\Delta_{i,j}\defeq \lambda_i - \lambda_j$ in the spectrum of an
unknown Hamiltonian without requiring entanglement with an external qubit.
While these gaps can be inferred directly using SMC, not all gaps will be self consistent.
It is therefore easier to learn the eigenvalues, which are unconstrained, than it is to
impose the appropriate constraints on the gaps.

Importantly, the RGE likelihood function is highly degenerate, as \autoref{eq:rge-like}
only depends on $\{\Delta_{i,j}\}$ and not on the eigenvalues of interest.
These degeneracies can be included analytically, so that we can verify
the estimates obtained by structured filtering.
It is worth noting that in cases where the eigenvalues are randomly distributed then
with high probability there will only exist two degenerate orderings and all other orderings 
can in principle be resolved by solving the turnpike problem.  Despite this, many approximate degeneracies are likely
to occur in such settings and these can be just as hazardous to learning as exact degeneracies.  As such, there is 
still a major need for structured filtering even in unstructured gap estimation problems.

As a final point, in practice it is likely that experimental data will be provided for RGE in batches if processing time for each update is long relative to data acquisition time.  We deal with this by assuming the data is acquired using $n_{\rm meas}$ measurements and the resulting distribution of $0$ or $1$ measurements is given by a binomial distribution with $p={\rm Pr}(0|H;t)$.  In practice, we take $3$ measurements in our numerical experiments.

\begin{figure}[t!]
    \subfloat[Prior]{
        \includegraphics[width=0.4\linewidth]{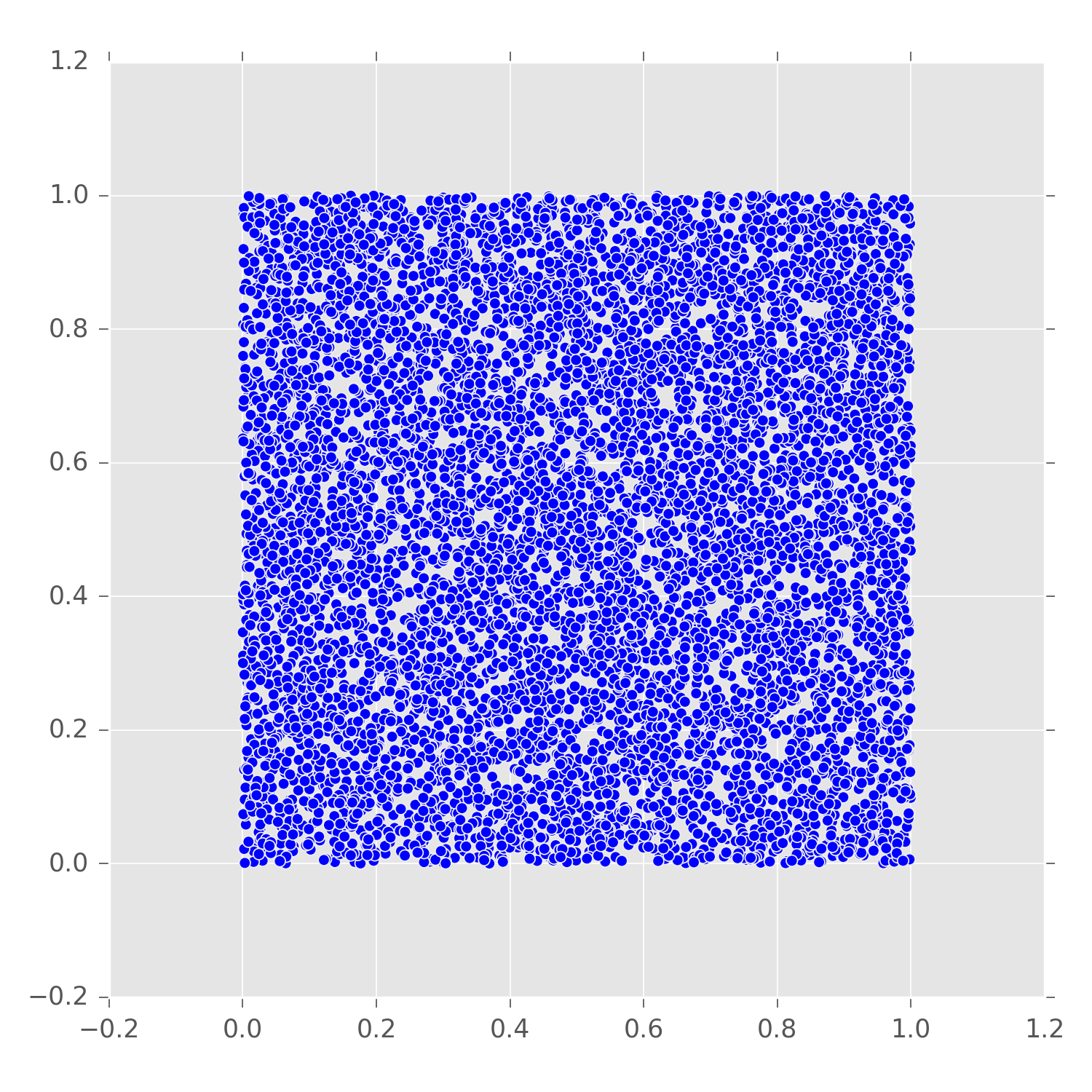}
    }
    \subfloat[After 100 updates]{
        \includegraphics[width=0.4\linewidth]{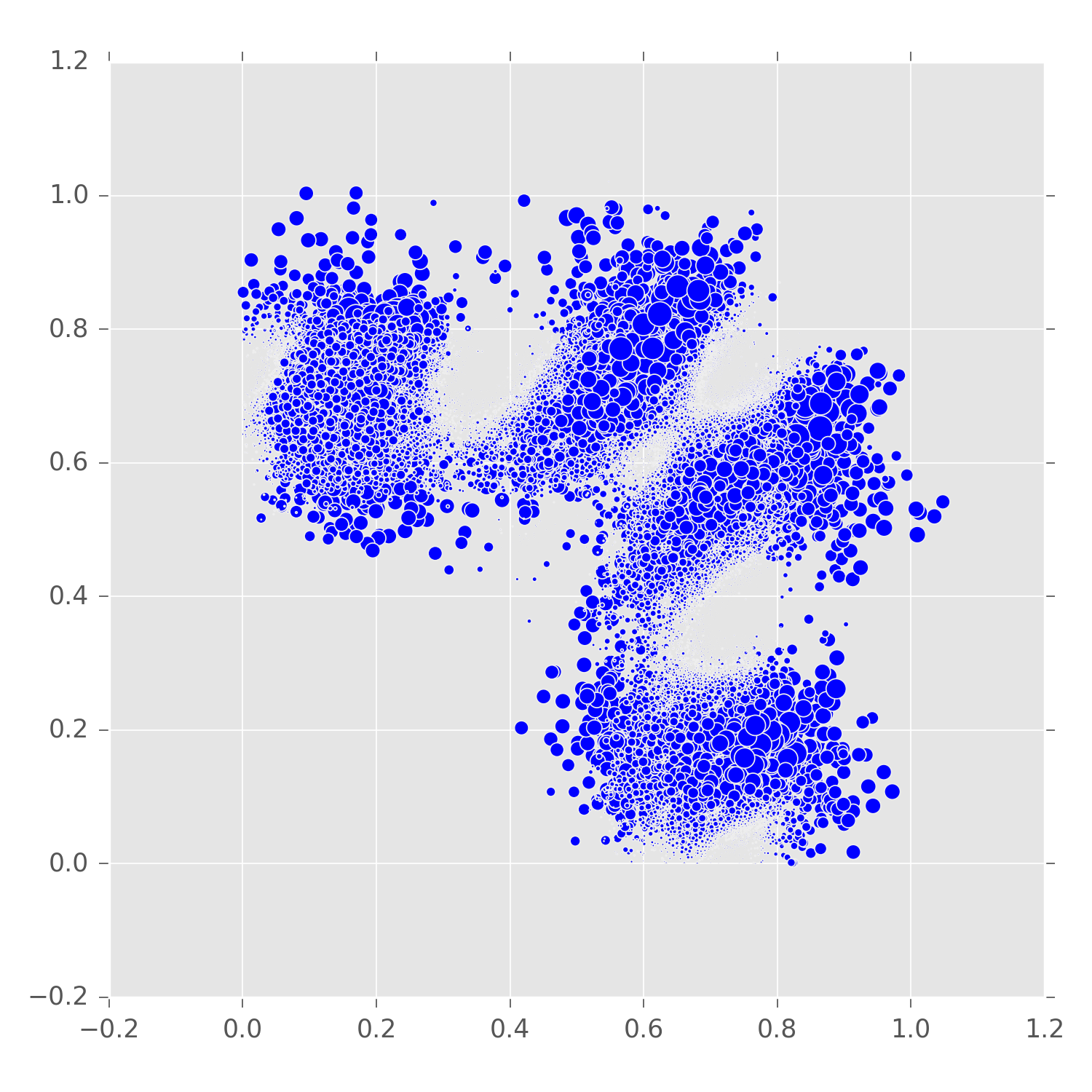}
    }

    \subfloat[After 150 updates]{
        \includegraphics[width=0.4\linewidth]{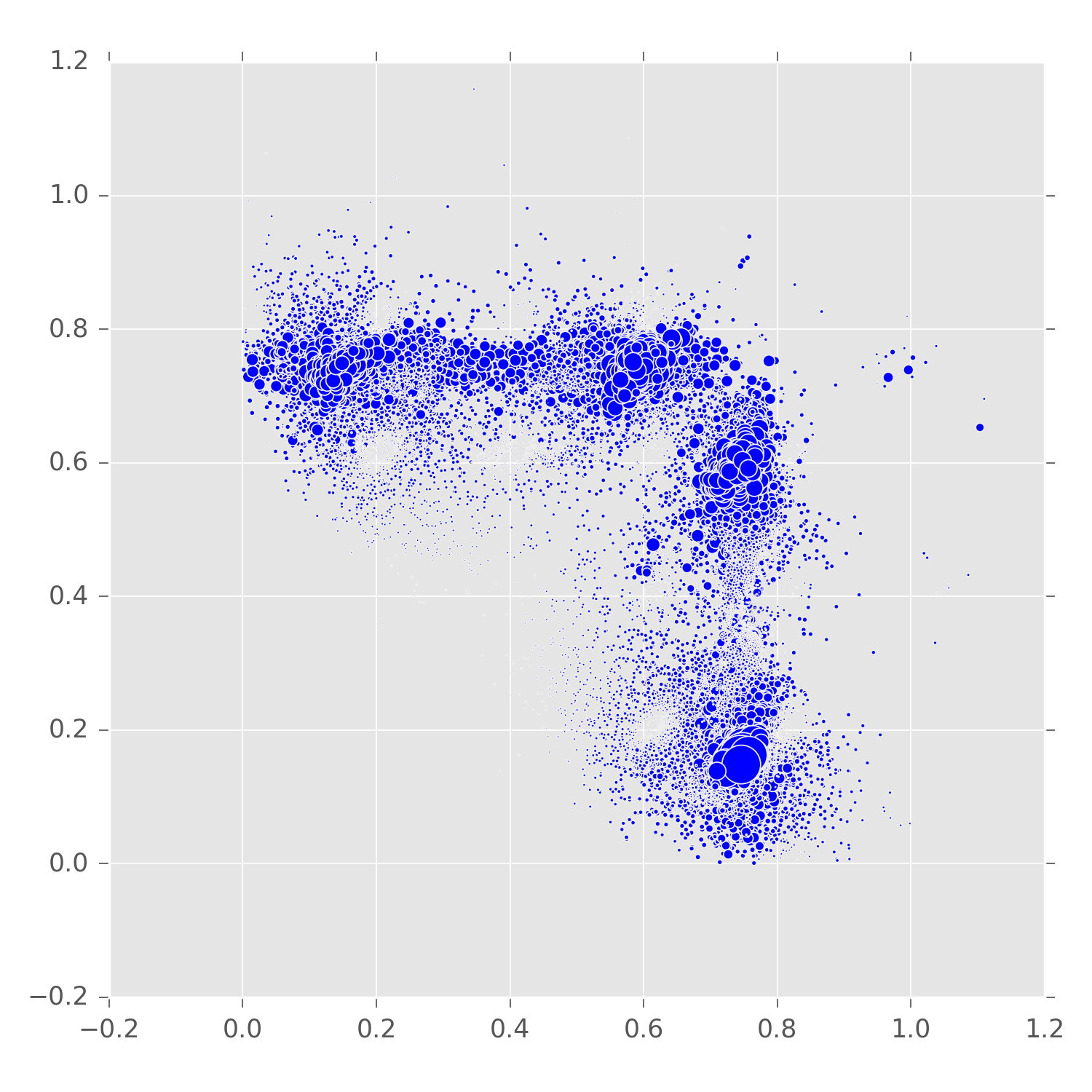}
    }
    \subfloat[After 250 updates]{
        \includegraphics[width=0.4\linewidth]{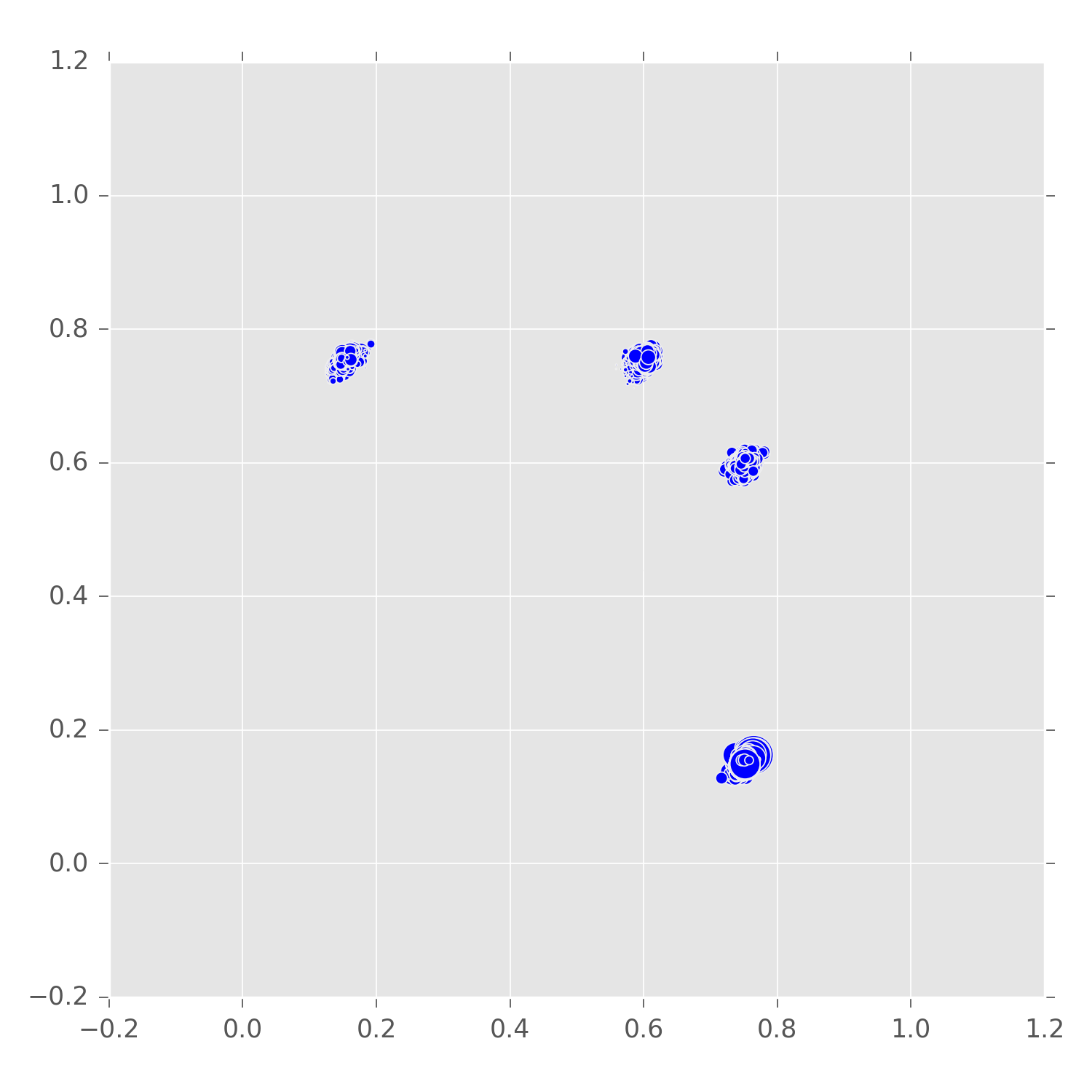}
    }
    \caption{\label{fig:example}
        Posterior distributions for using structured filtering to learn potential eigenvalues for randomized gap estimation with three eigenvalues where one is without loss of generality taken to be $0$.  The true model has unknown eigenvalues $(0.75, 0.15)$ and the symmetries of the problem imply the final posterior should ideally be concentrated at $(0.75,0.15), (0.15,0.75), (0.6,0.75), (0.75,0.6)$. The floor threshold was set to $0.1$ and champion threshold was set to $20$.  For this data and the graph was set to have maximum degree $3$ and maximum depth $4$.  Liu--West resampling with $a=0.98$ was used to cluster each cluster in the posterior and a minimum of $1000$ particles was assigned to each cluster.
        A video showing the operation of our algorithm is provided in the supplementary material, or online at
        \href{https://goo.gl/4NKYaX}{{\sffamily goo.gl/4NKYaX}}.
    }
\end{figure}

\begin{figure}
    \subfloat[Region estimate after 200 updates]{
        \label{subfig:example-metadata-region-est}
        \includegraphics[width=0.45\linewidth]{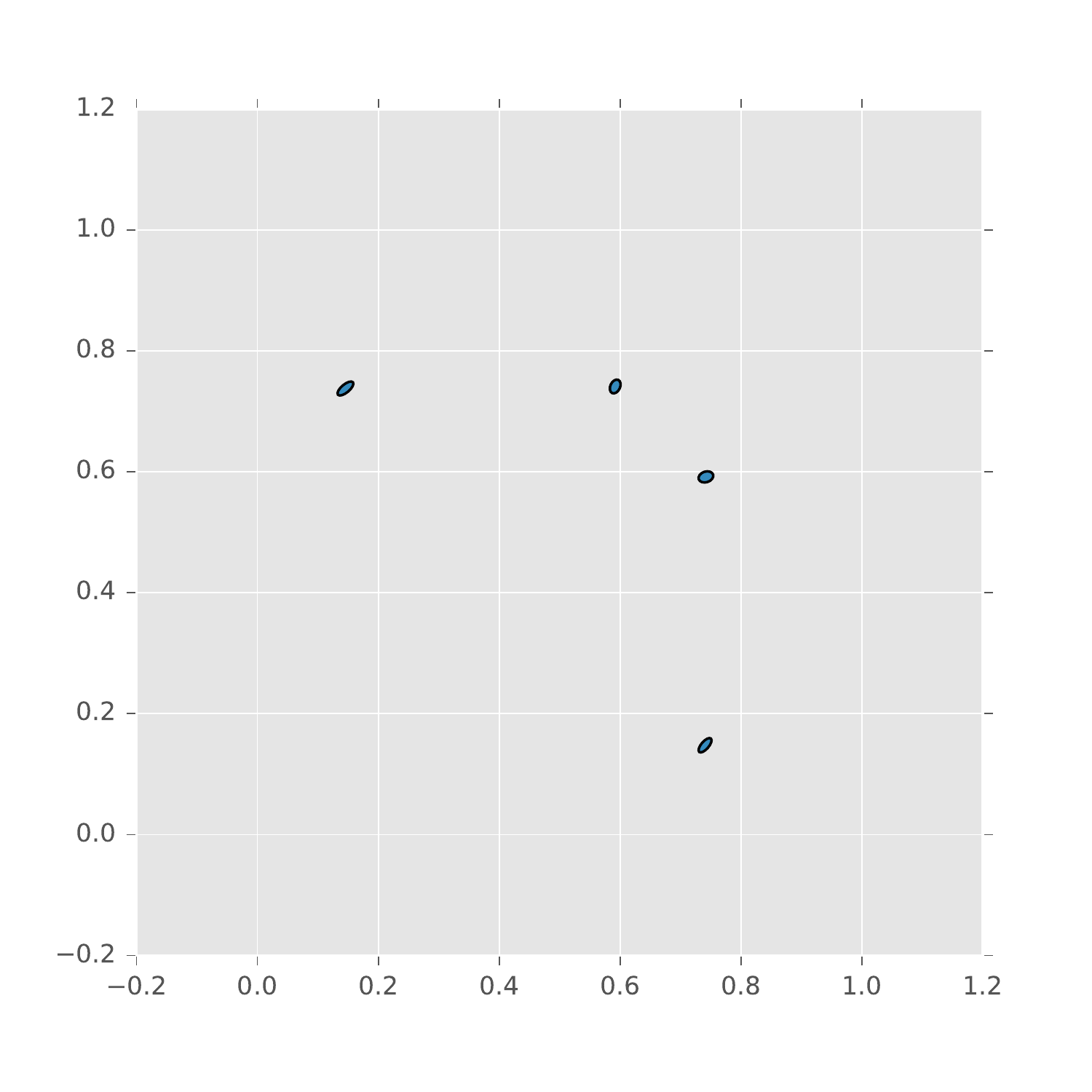}
    }
    \subfloat[Structural tree after 300 updates]{
        \label{subfig:example-metadata-final-tree}
        \includegraphics[width=0.45\linewidth]{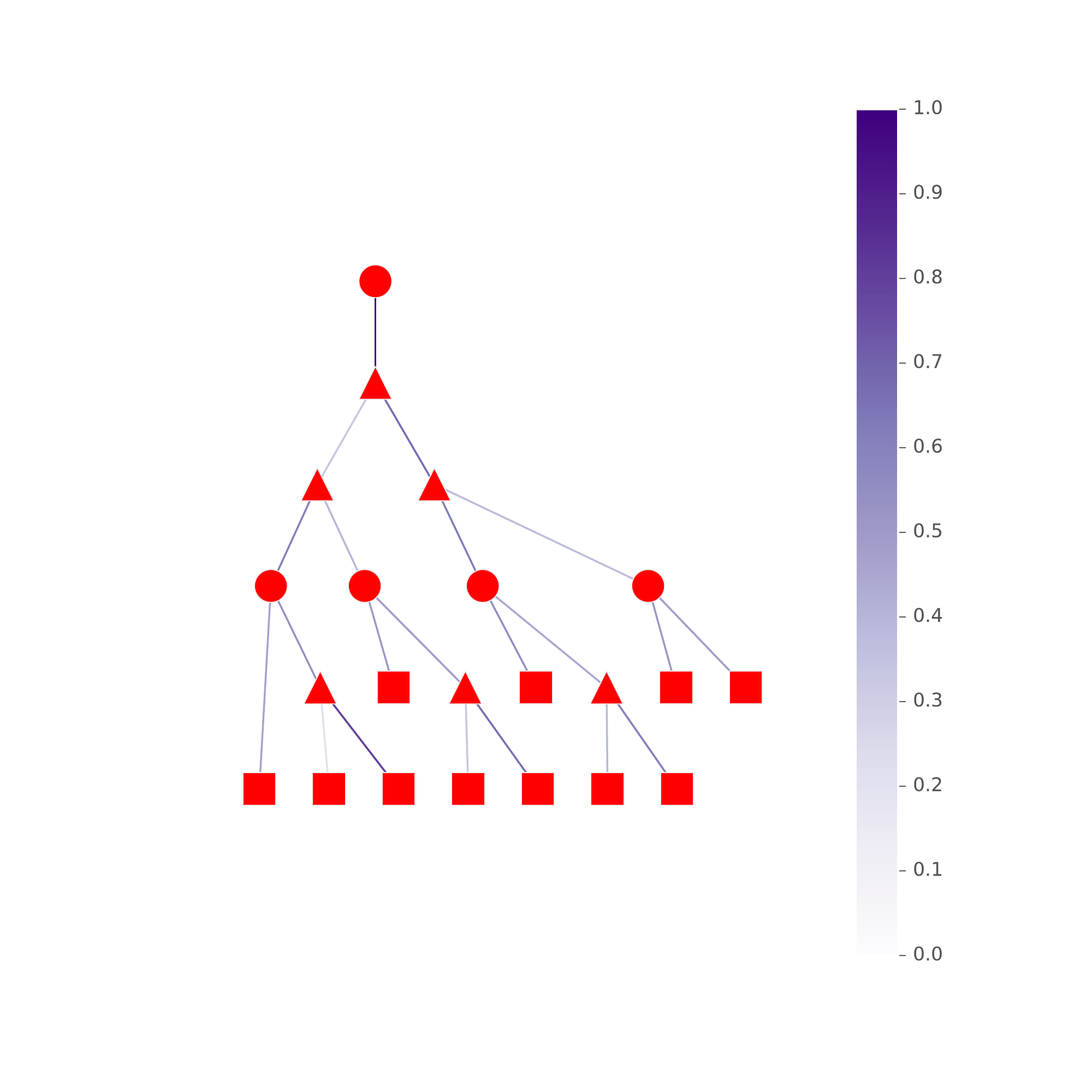}
    }
    \caption{\label{fig:example-metadata}
        (a) Region estimate and (b) structural tree for the example shown in
        \autoref{fig:example}.  Note the three mixture nodes that directly
        descend from the root in subfigure (b) indicate that the
        posterior consists of at least $4$ clusters, as each of the four
        children of these mixture nodes are decision nodes representing
        at least one cluster.
    }
\end{figure}

There are many ways that we could pick the experimental parameter $t$.  The best method is to choose $\theta$ to minimize the Bayes risk using a numerically optimized strategy.  This approach, while highly successful, is computationally expensive~\cite{granade_robust_2012}.  Another approach is to use a heuristic known as the \emph{particle guess heuristic} (PGH)
\cite{wiebe2014hamiltonian}.  The PGH, which is appropriate for periodic likelihoods like~\autoref{eq:rge-like}, guesses $t\propto 1/\sigma$ where $\sigma$ is an estimate of the uncertainty in the posterior distribution.  Here we approximate this by drawing unique $\vec{x}_1$ and $\vec{x}_2$ from the prior distribution and take 
\begin{equation}
    t = \frac{1}{\|\vec{x}_1 -\vec{x}_2\|_2^2}.\label{eq:pgh}
\end{equation}
The PGH is known to lead to asymptotically optimal scaling for phase estimation and slight variants of this have been shown to come close to saturating the Bayesian Cram\'er-Rao bound under the assumption of a Gaussian prior; however, less is known about its performance in multi--modal settings.

While this sampling procedure is straight forward for unstructured filtering, it is slightly more complicated in structured resampling.  This is because the structured filter is simultaneously entertaining a number of different hypotheses about the clusters that compose the posterior distribution.
We avoid this problem by computing the Bayes factors for each such hypothesis supported by the structured filter and only apply the PGH on the hypothesis that has the smallest Bayes factor.
That is, we purposely choose the experimental time based on the least probable explanation for the data.  We make this choice because it is likely to be conservative and also because it demonstrates the algorithms ability to learn despite being provided with inferior data.  The samples are then drawn from the remaining filter nodes according to their weights as per the PGH.

We first benchmark the performance of structured filtering as well as unstructured filtering for estimating the gaps in a three-level system (qutrit) where we have assumed without loss of generality that the lowest energy eigenvalue is $\lambda_0=0$.  We then define the eigenvalue gaps to be $\Delta_{i,j} = \lambda_j -\lambda_i$.  For convenience, let us assume that $\lambda_2 \ge \lambda_1$.  The three gaps in the problem are then $\Delta_{0,1}, \Delta_{1,2}, \Delta_{2,0}$.  The gaps that are learned can specify, up to an additive constant, the largest energy eigenvalue i.e. $\max\{\lambda_1, \lambda_2\} = \Delta_{0,2} = \Delta_{0,1} + \Delta_{1,2}$.  They cannot, however, uniquely give the first excited state.  This is because both $\lambda_1= \Delta_{0,1}$ and $\lambda_1=\Delta_{1,2}$ are consistent with the data.  There is also an obvious permutation symmetry in this problem wherein $\lambda_1 \leftrightarrow \lambda_2$ leaves the likelihood invariant.  In order to address these degeneracies in our
assessment of the algorithm, we define a canonical ordering of the eigenvalues and assignment of the gaps that in effect removes the four-fold degeneracy.  We then compute the \emph{canonical quadratic loss} as 
\begin{equation}
\mathcal{L}_{\rm can}:=\sum_i\left(w_i \|\vec{\lambda}_{{\rm canonical},i} - \vec{\lambda}_{\rm true,canonical}\|_2^2\right),
\end{equation}
 which we use as our figure of merit for the inference.  Here each $w_i$ is a particle weight taken from the clustered posterior distribution.

The multi-modal nature of the posterior distribution can be seen in~\autoref{fig:example}, wherein we consider a structured posterior that consists of a choice between $1$ and $2$ clusters at each splitting, with $d_{\max} = 4$, such that
our algorithm includes the optimal number of clusters $4$ as a possibility.  The figure provides the posterior distribution of \emph{every particle} in the filter wherein the size of the particle denotes the weight assigned to it in the filter node and all preceding mixture and decision nodes in the structure graph.  See supplementary information for an animation of the complete inference process.  We observe that after only $100$ updates, the structuring algorithm has clearly identified the structure of the ideal posterior, although it has not yet conclusively learned the number of clusters.  After $150$ updates we observe peaks in the probability density begin to congeal around the true eigenvalues (in particular near ($0.75,0.15$)).

At $250$ updates we see that the structured filtering algorithm correctly identifies all $4$ of the equivalent degenerate solution for the eigenvalues.  The algorithm further recognizes that the $4$ cluster model for the data is by far the best model for the posterior distribution.  This is signnificant because we make no apriori assumption that the ideal posterior is composed of $4$ clusters.  Furthermore, apart from giving an accurate point estimate of the eigenvalues, we obtain a posterior distribution from which uncertainties in these four estimations can be gleaned.  This information is of great value in experiments because principled estimates of uncertainty are often hard to find.
By contrast, our algorithm outputs them by default.

In addition to outputting the posterior distribution, structured filtering can output $\alpha$--credible regions and also structural information about the posterior distribution.
\autoref{subfig:example-metadata-region-est} provides an $\alpha=0.95$ credible estimate of a region where the true eigenvalues can be found.  Notably, because this region estimate does not output a single pair of eigenvalues its interpretation remains consistent regardless whether structured or unstructured filtering is used.  The region estimate also notably overlaps with the four modes of the ideal posterior distribution in this example.

\autoref{subfig:example-metadata-final-tree} gives the structure graph for the posterior after $300$ updates.  The three mixture nodes that directly descent from the root node reveal that the algorithm has completely excluded $1,2$ and $3$ cluster models as viable explanations for the data. This structure again was not imposed upon the posterior, nor was it directly included via our splitting rule.  Instead this structure emerged through the course of the $300$ updates by following the rules set out by structured filtering for both growing as well as pruning the structure tree.
This shows that structured filtering is not only able to learn patterns present in a complex posterior without substantial coaxing by the user, but also convey these inferred structures in a concise human-parsable format.
Additionally, by tracing through the structure graph from the root node, we can see that a four cluster (or approximately four cluster)
model is preferred in each branch considered.
In particular, each of the four children of the mixture nodes closest to the root are decision nodes
that prefer single-cluster explanations of the posterior.
This illustrates that structured filtering is capable of more than just inferring the structure of the posterior distribution: it is also capable of reporting it in a human readable format.

\begin{figure}
    \begin{center}
    \begin{minipage}{0.49\linewidth}
            \includegraphics[width=\textwidth]{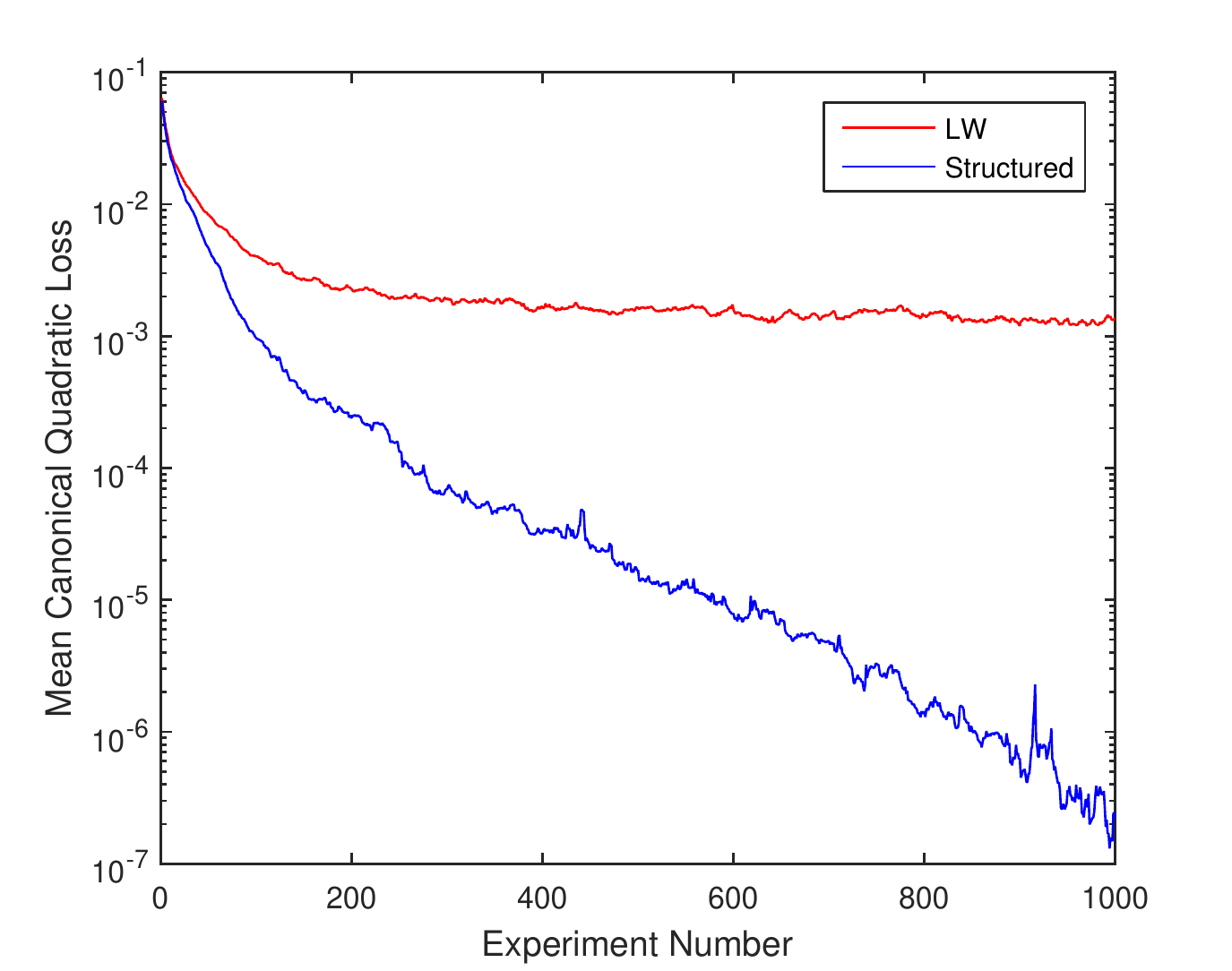}
    \end{minipage}
    \hspace{1mm}
    \begin{minipage}{0.49\linewidth}
            \includegraphics[width=\textwidth]{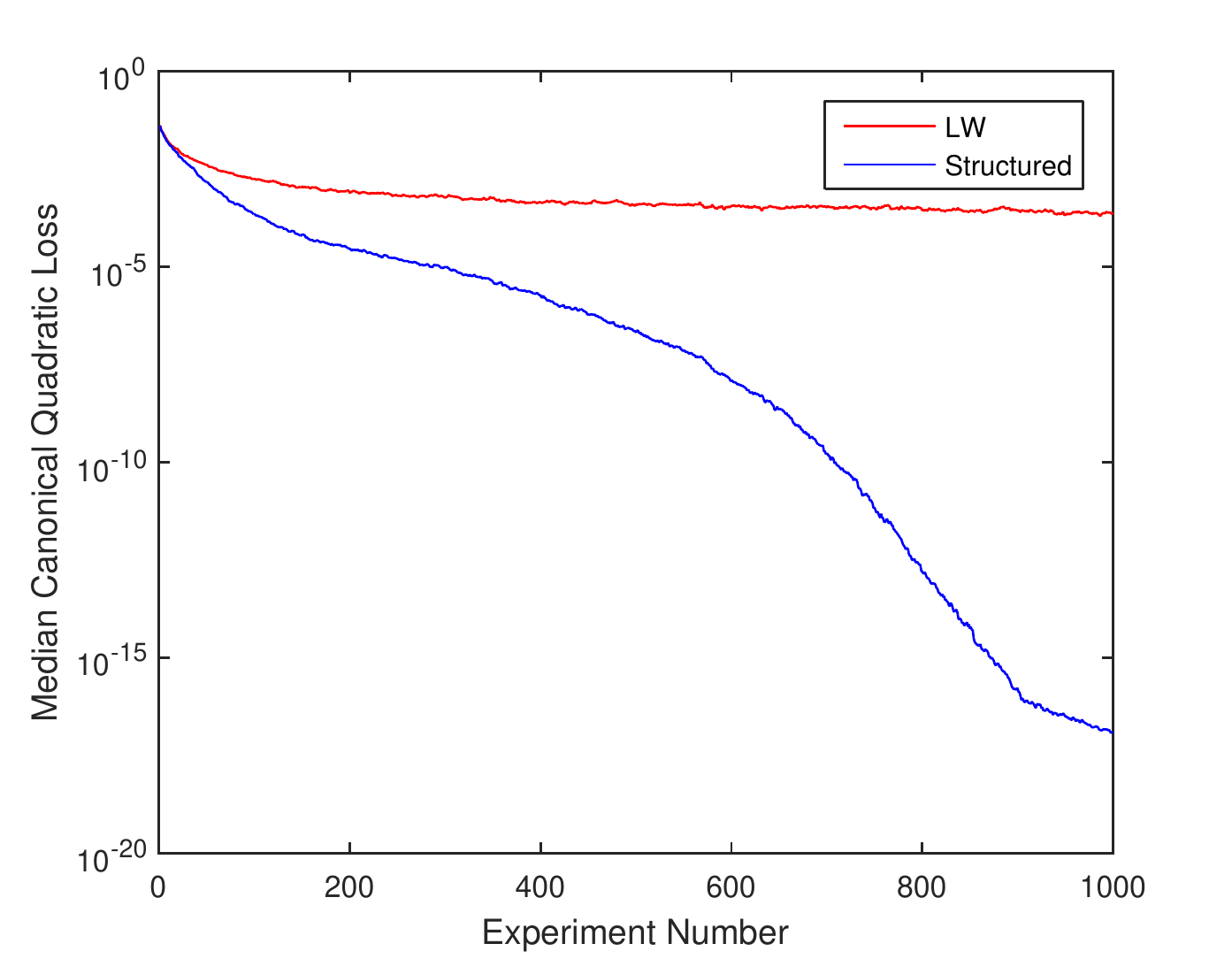}
    \end{minipage}\\
    \begin{minipage}{\linewidth}
        \parametertable{
            $a$ & $0.98$ & $d_{\max}$ & $4$ \\
            $N_{\rm part}$ & $8000$ & $n_{\rm clusters}$ & $\{1, 2\}$ \\
            $h$ & $0.5$ & $w_{\rm floor}$ & $0.1$ \\
            $n_{\rm meas}$ & $3$ & $K_{\rm champ}$ & $2000$ \\
            & & $\min n_{\rm part}$ & $1000$\\
        }
    \end{minipage}

    \end{center}
    \caption{
        \label{fig:meanRGE}
        Comparison of Liu-West resampling and structured filtering
        for randomized gap estimation (RGE).  Means and medians are computed using $1000$ random RGE instances, with $\lambda_1$ and $\lambda_2$ sampled from the initial prior which is uniform on $[0,1]^2$.
    }
\end{figure}

\autoref{fig:meanRGE} shows that this degeneracy in the likelihood function causes LW to fail in both the mean and median $\mathcal{L}_{\rm can}$ after roughly $200$ experiments where it saturates at a value on the order of $10^{-3}$.    This is to be expected because the unimodality assumption implicitly made in Liu--West resampling will generically fail here just as it did in~\autoref{fig:lw-resampling-failure}.  

Structured filtering, on the other hand, can learn the correct eigenvalues within numerical precision with high probability after only $1000$ experiments, and further we see clear evidence of exponential scaling of the mean canonical loss.  This implies that even the worst case behavior of the algorithm is not pathological.  We find from linear regression that after $400$ experiments ${\rm mean}(\mathcal{L}_{\rm can}) \approx e^{-0.0083 n -6.7}$.   The data for the median does not demonstrate a single clear asymptotic scaling.  The appearance of two different time scales for the problem likely correspond to the timescales needed to distinguish the two degenerate possibilities for the eigenvalues (i.e. solve the turnpike problem on the gaps) and the timescale needed to refine this knowledge once a degenerate set of solutions to the turnpike problem is found.

This clearly shows that in cases where we have a multi-modal posterior that structured filtering can succeed where the gold-standard particle filtering methods used in quantum experiments fail.  
It succeeds here because structured filtering is capable of recognizing the multi-modal nature of the posterior and adjust the particle filter accordingly.  LW cannot succeed because it always assumes a unimodal posterior and here the ideal posterior has $4$ modes.
For this reason, we expect structured filtering to be a broadly applicable method capable of dealing both with the degeneracies that appear in randomized gap estimation and also approximate degeneracies that appear in other applications.

\section{Application to Collapse-Free Phase Estimation}
\label{sec:phase-est}

Phase estimation has become a ubiquitous algorithm in quantum computing because of its ability to learn eigenvalues of a unitary using quadratically fewer queries to that unitary.  Of the many variants of phase estimation, iterative phase estimation has gained in popularity owing to the fact that it does not require a large qubit register to store the estimated eigenvalues~\cite{kitaev2002classical}.  It works by replacing the quantum interference step used in traditional phase estimation with an adaptive inference algorithm to learn the most likely phase given a set of measurements.  This approach requires a number of queries to the blackbox that is optimal to within a small multiplicative factor and is thus also the preferred technique if speed is also an issue.

Both iterative and traditional phase estimation  require long sequences of gates in order to learn the eigenstate, which is perhaps the biggest reason why most simulation experiments eschew this approach at present~\cite{o2016scalable}.  These long sequences arise because long experimental times are needed to unambiguously collapse the state onto an eigenvalue with a small number of experiments according to energy/time uncertainty.   One approach to address is this issue is to shift the burden of collapsing the state from the quantum computer to the classical inference algorithm by applying phase estimation on a fresh copy of the initial state with each experiment.  We call this approach \emph{collapse-free phase estimation} as it does not collapse the quantum state onto an eigenstate of the Hamiltonian.

A challenge facing collapse-free phase estimation is that signal is received from multiple eigenvalues when the input state is a superposition of different eigenstates.  Thus any estimate of a single eigenvalue for the distribution will have to disambiguate data that comes from distinct eigenvalues.  One approach to dealing with this issue is to mimic wave function collapse by implicitly introducing biases against data that is more likely to come from other eigenvalues.
This approach is taken by~\citet{PhysRevLett.117.010503} and by \citet{santagati2016quantum}, wherein a Bayesian form of phase estimation is used to introduce such biases through applying a unimodal approximation to the posterior.

While unimodal approaches do not utterly fail here, they are not expected to perform as well because of the multi-modal nature of the ideal posterior if multiple eigenvalues are present.  To examine this further, let us first define
\begin{equation}
\ket{\psi} = \sum_{j} a_j \ket{E_j},
\end{equation}
where $E_j$ is the $j^{\rm th}$ eigenvalue of a Hamiltonian $H$.
Given such an initial state, the likelihood of measuring ``zero'' in the two outcome phase estimation experiment for $e^{-iHt}$ given experimental parameters $t$ and $\theta$ is
\begin{equation}
P(0|\{E_j\};\{a_j\},t,\theta)=\sum_j |a_j|^2 \cos^2([E_j -\theta]t/2).\label{eq:mixcosP}
\end{equation}
In collapse-free phase estimation, we do not know how many eigenvalues there are in the support of $\ket{\psi}$.  We also do not want to track them explicitly because there are exponentially many in the worst case.  Instead, we use the following likelihood function to model the experiment:

\begin{equation}
P(0|E;t,\theta)=(1-h) \cos^2([E -\theta]t/2)+ h,
\end{equation}
where $h\in [0,1]$ is a hedging parameter that is used to combat overconfidence in an update.  This model in essence assumes that all the data arises from a single eigenvalue, and aims to estimate the most likely single $E$ given data that arises from~\autoref{eq:mixcosP}.

We similarly use~\autoref{eq:pgh} to choose the experimental times and take $\theta=\vec{x}_2$ in the PGH.  Since this heuristic is known to provide asymptotically optimal scaling in cases where only one eigenvalue is present~\cite{wiebe2014hamiltonian}, we anticipate that it will provide similar advantages here as well.

We consider two eigenvalues with uniformly distributed values between $0$ and $1$ and attempt to learn one of the two eigenvalues.  If the two eigenvalues are $E_1$ and $E_2$ then the loss function we consider is:
\begin{equation}
\mathcal{L}_{\min}  = \min_{E\in \{E_1,E_2\}}\sum_i w_i (E-x_{i})^2,
\end{equation}
where $x_i$ is the position of particle $i$ each of which corresponds to an eigenvalue of $H$.  We then consider the mean and the median over $1000$ randomly pairs of eigenvalues, where we pick the worst case scenario of $a_1 = a_2 =1/\sqrt{2}$.

\begin{figure}
    \begin{center}
        \begin{minipage}{0.49\linewidth}
                \includegraphics[width=\textwidth]{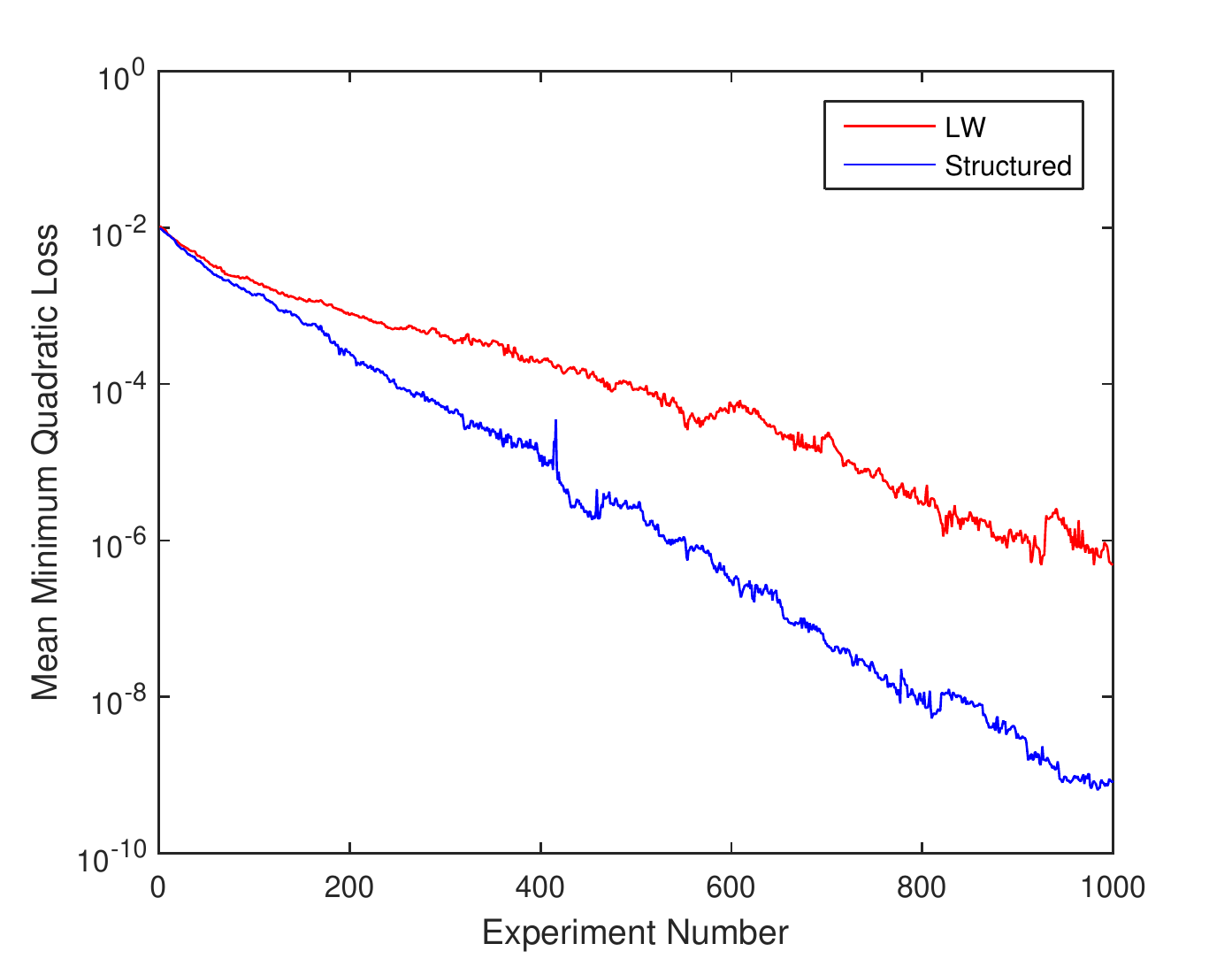}
        \end{minipage}
        \hspace{1mm}
        \begin{minipage}{0.49\linewidth}
            \includegraphics[width=\textwidth]{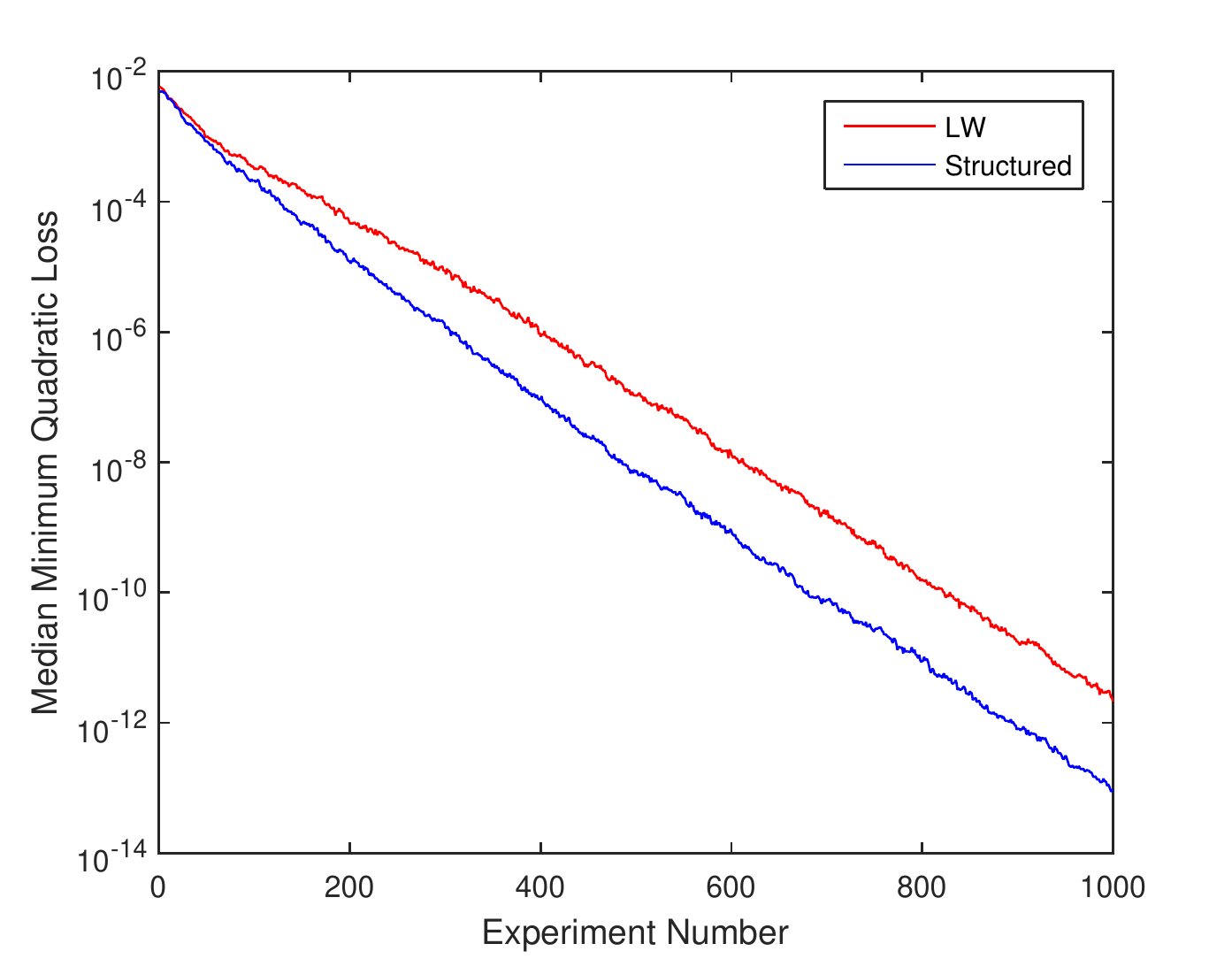}
        \end{minipage}
        \\
        \begin{minipage}{\linewidth}
           \parametertable{
                $a$ & $0.98$ & $d_{\max}$ & $6$ \\
                $N_{\rm part}$ & $12000$ & $n_{\rm clusters}$ & $\{1, 2\}$ \\
                $h$ & $0.5$ & $w_{\rm floor}$ & $0.1$ \\
                & & $K_{\rm champ}$ & $2000$\\
                & & $\min n_{\rm part}$ & $1000$\\
            }
        \end{minipage}
    \end{center}
    \caption{
        \label{fig:meanMixCos}
        Comparison of Liu-West resampling and structured filtering
        for collapse-free phase estimation with two true eigenvalues. 
    }
\end{figure}

We examine the performance of structured filtering and Liu--West resampling for this problem in~\autoref{fig:meanMixCos}.  Specifically we see that while both methods are capable of rapidly learning one of the eigenvalues, Liu--West resampling performs substantially worse.  This is not simply because structured filtering used more particles;  Liu--West resampling does not give significantly better results when we allow it more than $12000$ particles.  These differences are, unsurprisingly, most striking in the mean.  With Liu--West resampling linear regression shows that the mean minimum quadratic loss decays roughly as $\mathcal{L}_{\min} \approx[ 8.1\times 10^{-3}]  e^{-0.0095n}$ whereas the quadratic loss decays for structured filtering as $\mathcal{L}_{\min} \approx[ 6.7\times 10^{-3}]  e^{-0.0164n}$, where $n$ is the number of experiments.  This shows that the two asymptotic scalings differ polynomially, specifcially by a power of rougly $3/2$.  This shows that structured filtering can improve the performance of collapse-free phase estimation compared to Liu--West resampling, which is arguably the most powerful efficient inference method previously used for such problems.

\section{Conclusion}
\label{sec:conclusion}

Existing methods for parameter estimation in quantum systems are implicitly biased via structural assumptions about the posterior distribution.
Here we show that even the subtle biases introduced through filtering with Liu--West resampling
can catastrophically fail to estimate parameters for problems
where the experiments chosen are incapable of distinguishing between two equivalent hypotheses.  We address this by introducing a method
that can adaptively reason about the structure of the posterior distribution and break it up into clusters that individually are ammenable to Liu--West filtering or other methods that are appropriate for unimodal distributions.  Specifically we represent the algorithm's beliefs about the structure of the posterior as a weighted graph that describes the different possible clusterings for the posterior distribution, and allow it to invent new hypotheses or eliminate previous ones via a discrete set of graph manipulations.  By using this approach we grant structured filtering the ability to introspect on its own beliefs about the true model parameters.  This introspection is key to the success of our algorithm.

We note that by allowing the inference algorithm to adapt its assumptions about the structure of its current state of knowledge we allow Bayesian inference to succeed in problems, such as randomized gap estimation,
where existing leading methods fail.  It also manages to outperform its unstructured counterpart in learning eigenphases of unitary
operations in settings where the data comes from multiple eigenvalues.  This illustrates the power and broad applicability of the technique.

Looking forward, there are many ways that these methods can be built upon.  Firstly, while our approach does an excellent job of approximating
the posterior distribution it does a less perfect job of approximating the regions of importance in the posterior distribution.  In particular, our rules
for restructuring the graph assume that regions of hypothesis space that have low probability have low importance; however, since Bayes' rule is additive
in logarithmic space, wild fluctuations can occur in the posterior probability.  Such fluctuations can be common in cases like collapse-free phase estimation
because the approximate likelihood does not match the true likelihood.  This means that it may be important for the structured filters to learn how to distribute
particles to accommodate data that is given low probability by the assumed likelihood function.

Furthermore, structured filtering can be viewed as an application of a family of techniques known as probabilistic programming.  Further work will be needed to see if
subsequent insights from probabilistic programming may yield even more powerful representations for the posterior distribution.

Finally, while we have provided a proof of principle that structured filtering can solve many of the problems plaguing SMC approximations in
physics, it remains to apply them experimentally.  It is our hope that these methods may prove to be of great use estimating
Hamiltonian parameters that have subtle influence on experimental likelihoods, such as those that appear in second order corrections to spectral line splittings.  Structured resampling, and approaches like it,
may finally have enough power and robustness to tackle such problems in an automated fashion relieving the experimentalist of much of the
burden of designing clever experiments to learn hard to measure quantities.


\begin{acknowledgments}
    CG was supported by the US Army Research Office grant numbers W911NF-14-1-0098 and W911NF-14-1-0103,
    and by the Australian Research Council Centre of Excellence for Engineered Quantum Systems. 
    We thank Sarah Kaiser and Chris Ferrie for helpful comments, and
    \citet{okabe_color_2002} for their suggestion of a colorblind-safe
    palette for figures.
\end{acknowledgments}

\bibliography{structured-filtering}

\onecolumngrid
\appendix

\section{Additional Numerical Experiments}
\label{apx:numerics}

While considerable experience has been built over the last few years studying how to optimally learn parameterizations of periodic likelihoods,
the meta parameters needed to allow structured filtering to succeed in these cases is not known.  Furthermore, it isn't clear how robust structured
filtering is to the choice of parameters.  Here we perform experiments to probe these issues for RGE.

\begin{figure}[t!]
    \begin{minipage}{0.49\linewidth}
        \includegraphics[width=\linewidth]{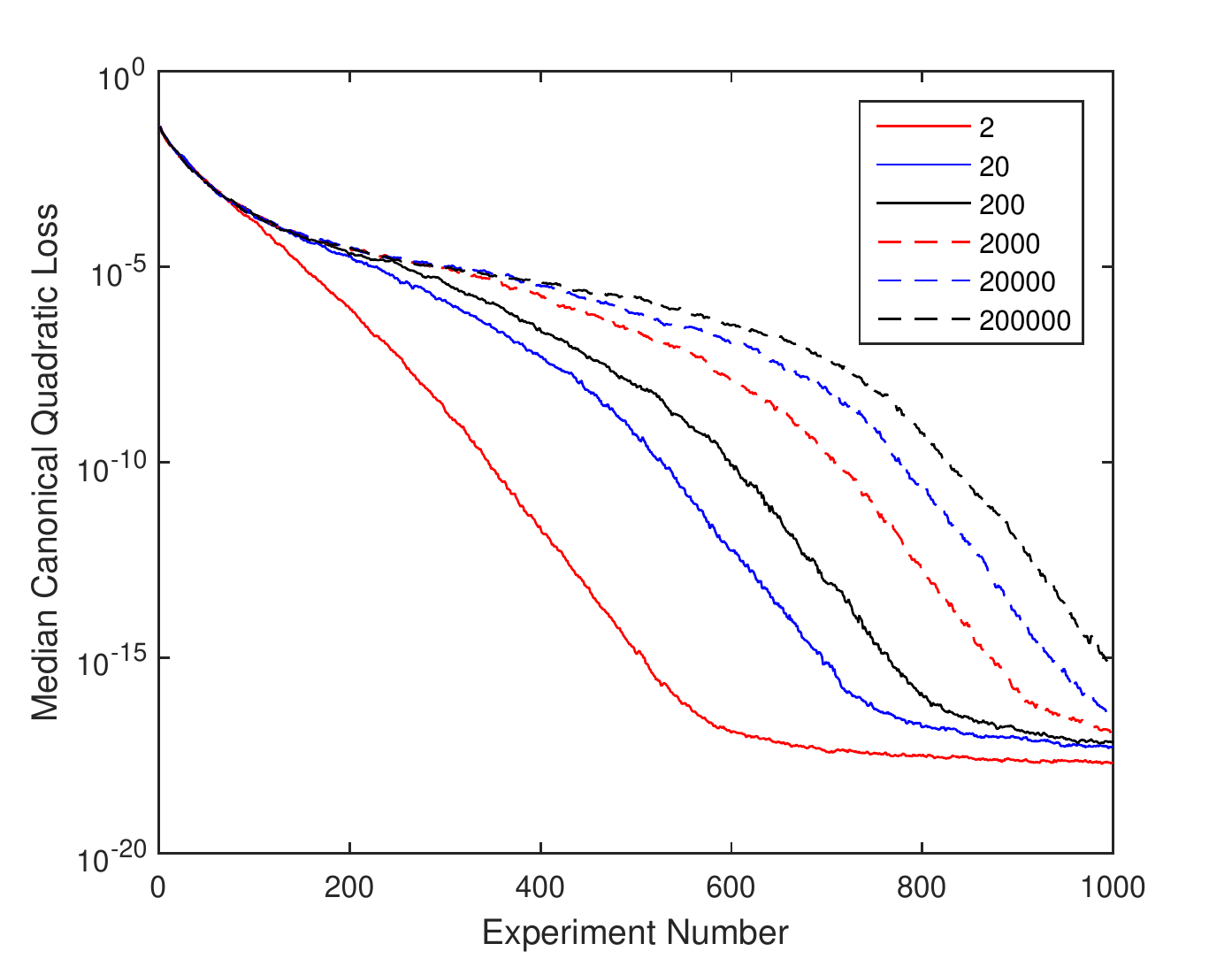}
    \end{minipage}
    \hspace{1mm}
    \begin{minipage}{0.49\linewidth}
        \includegraphics[width=\linewidth]{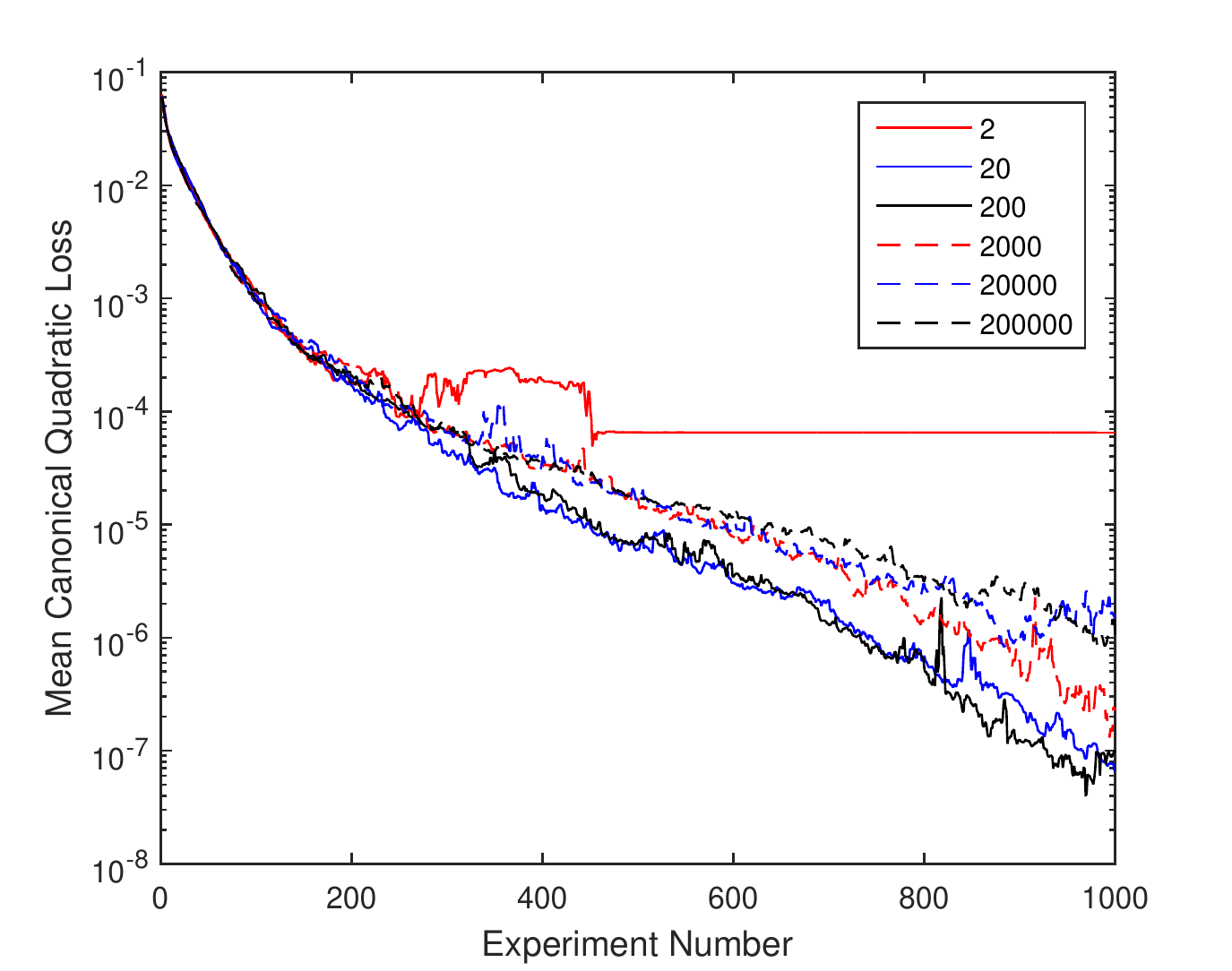}
    \end{minipage}

    \begin{minipage}{\linewidth}
        \parametertable{
            $a$ & $0.98$ & $d_{\max}$ & $4$ \\
            $N_{\rm part}$ & $8000$ & $n_{\rm clusters}$ & $2$ \\
            $h$ & $0.5$ & $w_{\rm floor}$ & $0.1$ \\
            $n_{\rm meas}$ & $3$ & $\min n_{\rm part}$ & $1000$ \\
        }
    \end{minipage}
    \caption{
        Median and mean canonical quadratic loss for RGE with $3$ eigenvalues as a function of the champion ratio. All structured parameters are applied
        at the root context.
    }\label{fig:champ}
\end{figure}

\autoref{fig:champ} provides numerical experiments that probe the mean and median performance of structured filtering for randomized gap estimation
as a function of $K_{\rm champ}$.  We see from the median data that as $K_{\rm champ}$ shrinks the performance of the algorithm improves.  This
improvement comes about in part because of a coupling with the version of the particle guess heuristic that we choose.  Since we guess experimental
times based on the worst covariances kept in the problem, choosing a high champion ratio can retain poor models which leads to less informative experiments
yielded by this variant of the PGH.  This is why we expect, and observe, that taking $K_{\rm champ}=2$ provides the best performance in the median.

The mean canonical loss given in~\autoref{fig:champ} tells a different story.  Since the mean is not a robust statistic, it is sensitive to rare instances
where the errors are much larger than the typical cases.  We see from this data that while the performance tends to improve as $K_{\rm champ}\rightarrow 20$, the
algorithm is much more likely to catastrophically fail for $K_{\rm champ}=2$.  Again this is expected because as $K_{\rm champ}$ goes to $1$ we expect that the probability
of falsely concluding that the most probable model is the correct one increases.  Thus the optimal choice of $K_{\rm champ}$ is analogous to trading off type $1$ and type $2$
errors in hypothesis testing.

For our numerics we wished to avoid catastrophic errors that could dominate the scaling of the mean so we chose $K_{\rm champ}=2000$.  While this value dramatically increases
the number of experiments needed to solve RGE problems in the median according to~\autoref{fig:champ} relative to $K_{\rm champ} =2$ or $20$.  In principle, these drawbacks may be addressable with improved guess heuristics for the experimental times here.  We leave a more thorough investigation of such questions for subsequent work.

\begin{figure}[t!]
    \begin{minipage}{0.49\linewidth}
        \includegraphics[width=\linewidth]{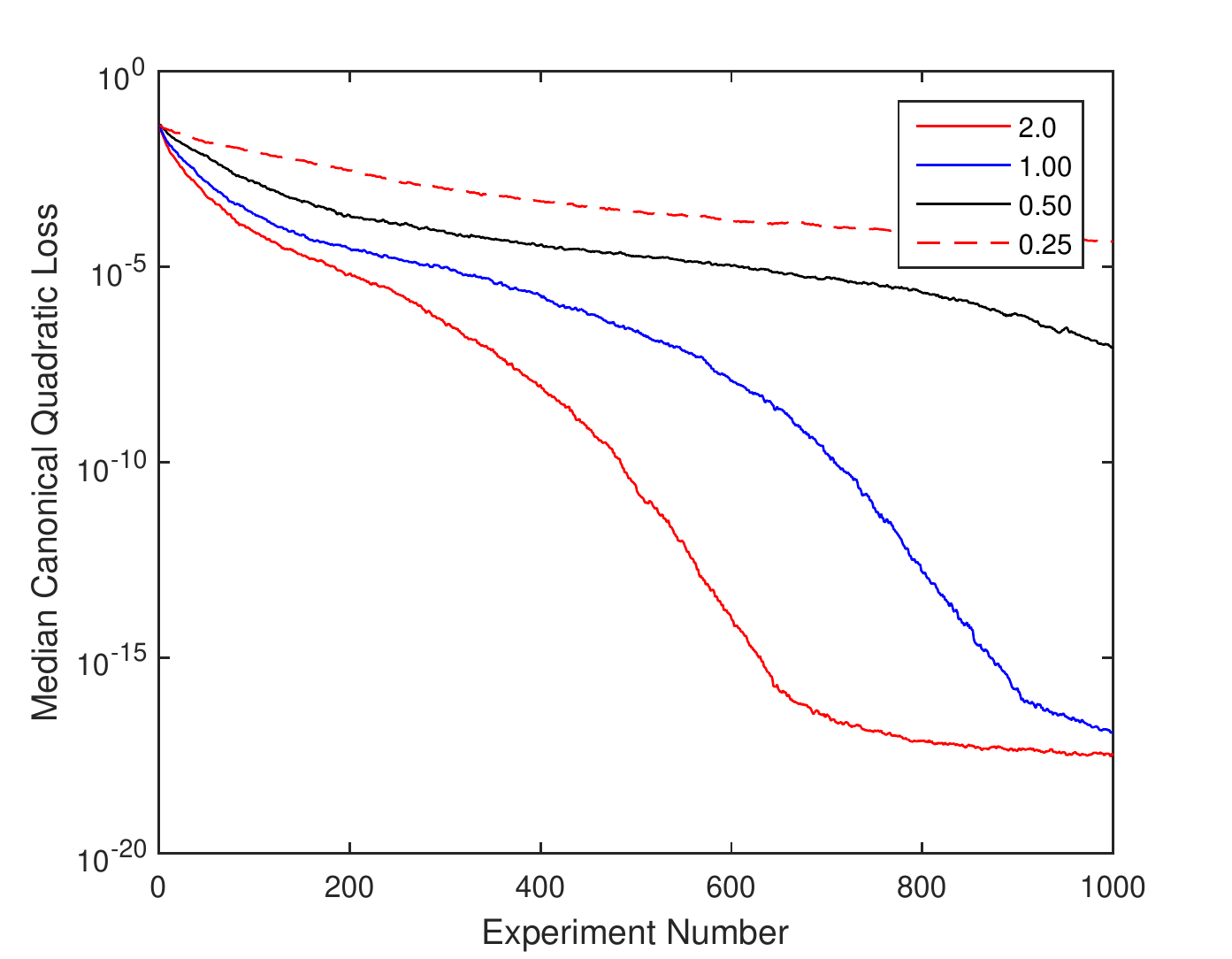}
    \end{minipage}
    \hspace{1mm}
    \begin{minipage}{0.49\linewidth}
        \includegraphics[width=\linewidth]{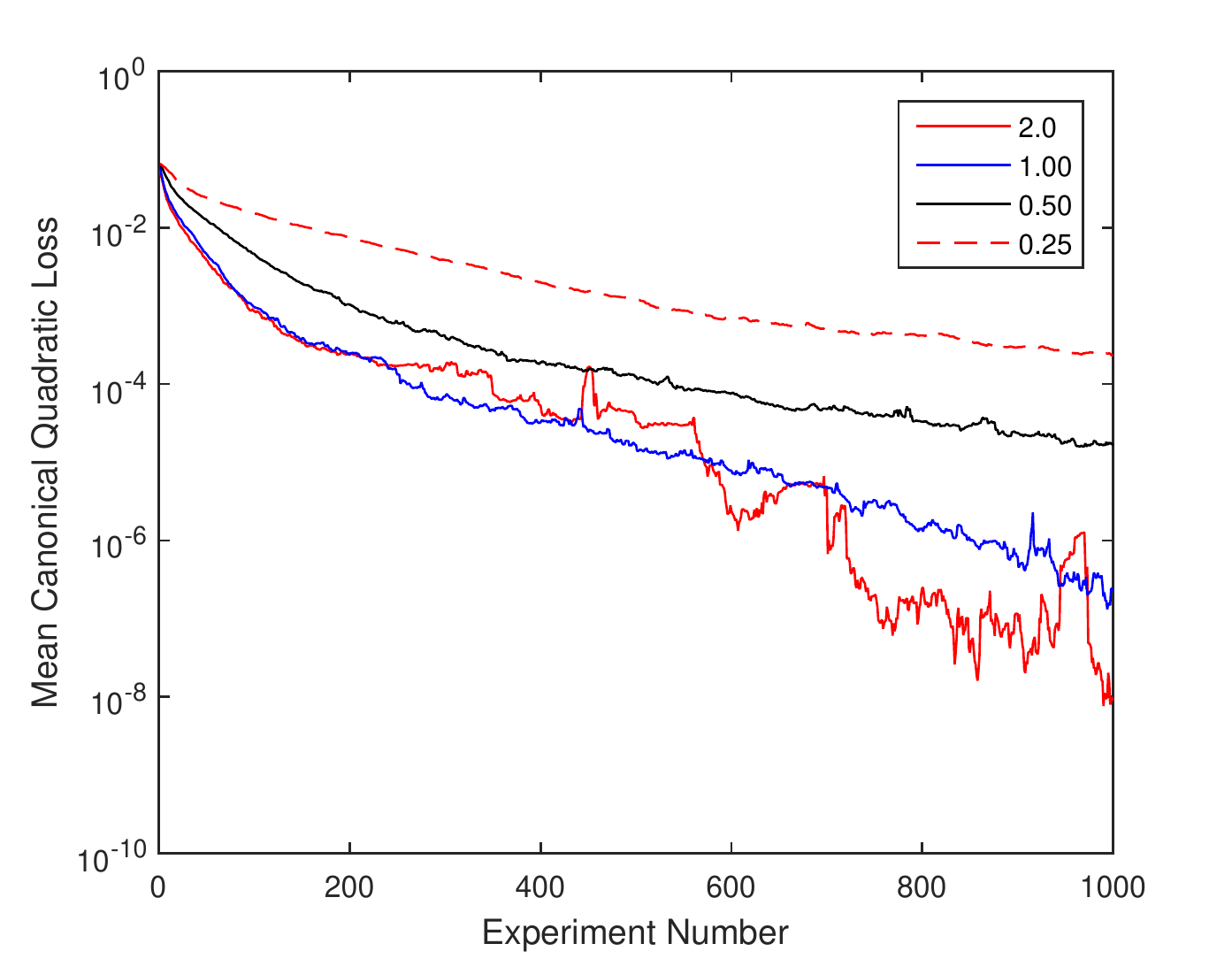}
    \end{minipage}
    \begin{minipage}{\linewidth}
        \parametertable{
            $a$ & $0.98$ & $d_{\max}$ & $4$\\
            $N_{\rm part}$ & $8000$ & $n_{\rm clusters}$ & $2$ \\
            $h$ & $0.5$ & $w_{\rm floor}$ & $0.1$ \\
            $n_{\rm meas}$ & $3$ &  $\min n_{\rm part}$ & $1000$ \\
            & & $K_{\rm champ}$ & $2000$\\
        }
    \end{minipage}
    \caption{
        Median and mean canonical quadratic loss for RGE with $3$ eigenvalues as a function of the constant for the PGH used.  A PGH constant of $1$, i.e. $t=1/\|x_{1} -x_{2}\|_2$ for $x_1$ and $x_2$ sampled from the prior, is used in the remainder of the paper. All structured parameters are applied
        at the root context.
        \label{fig:PGH}
    }
\end{figure}

\autoref{fig:PGH} addresses the question of how the experimental times should be guessed for RGE experiments using structured filtering.  Previous work, suggested that for alternative filtering strategies such as rejection filtering~\cite{zintchenko_randomized_2016}, a pgh constant of $2$ performs well.  We see similar results here, with a constant of $2$ outperforming the standard choice used in previous Hamiltonian learning work of $1$.  We see that this improved performance is exhibited in both the median and the mean here, although the mean performance for a constant of $1$ seems much more variable than it is for larger values of the constant.  While this is not necessarily a negative thing because it shows that the error is both low and dominated by a few outliers, it makes a Monte--Carlo estimations of the mean more expensive.  For this reason we retain the pgh constant of $1$ instead of $2$ and note that again these results can be optimized by choosing more intelligent experiments.

\begin{figure}[t!]
    \begin{minipage}{0.49\linewidth}
        \includegraphics[width=\linewidth]{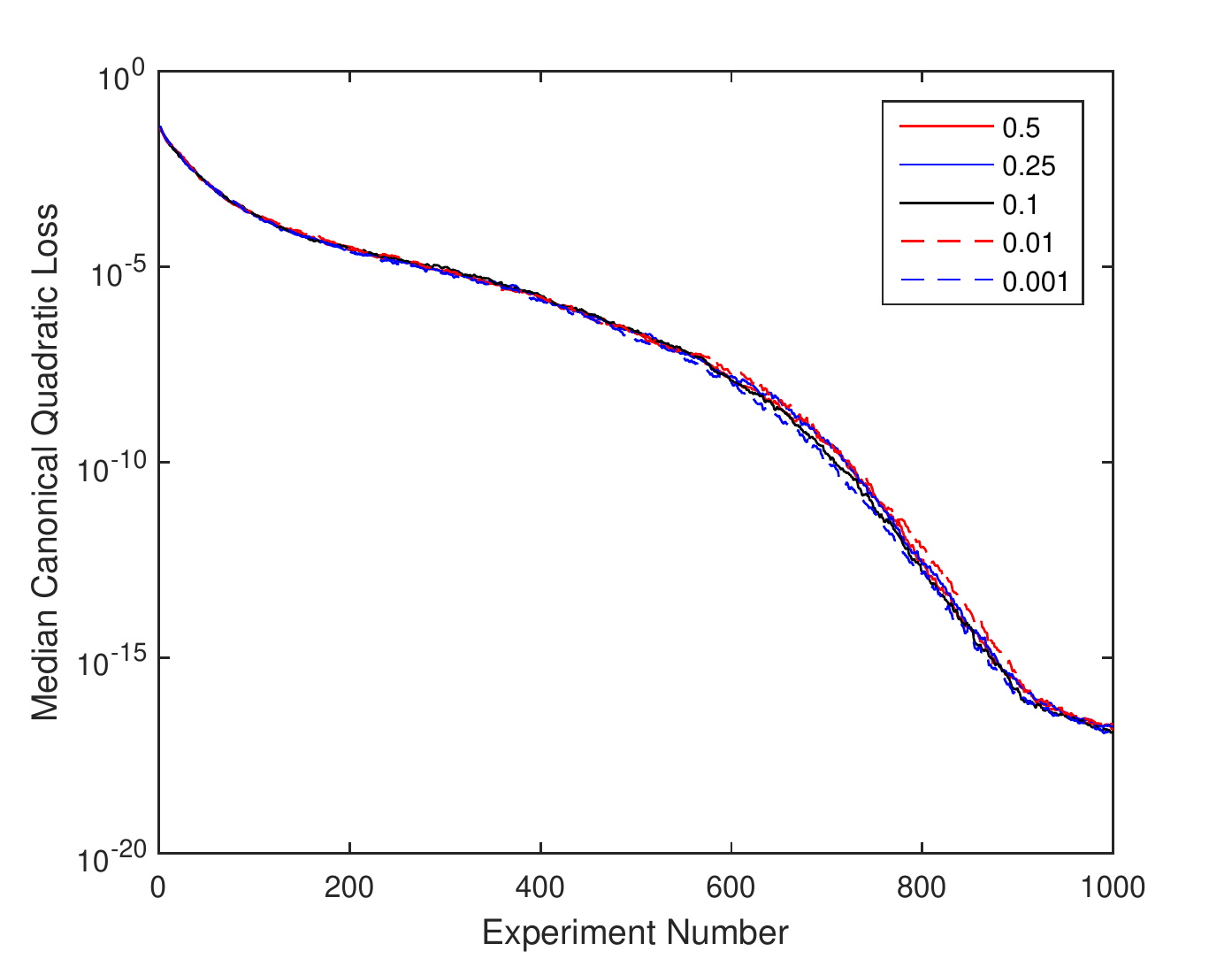}
    \end{minipage}
    \hspace{1mm}
    \begin{minipage}{0.49\linewidth}
        \includegraphics[width=\linewidth]{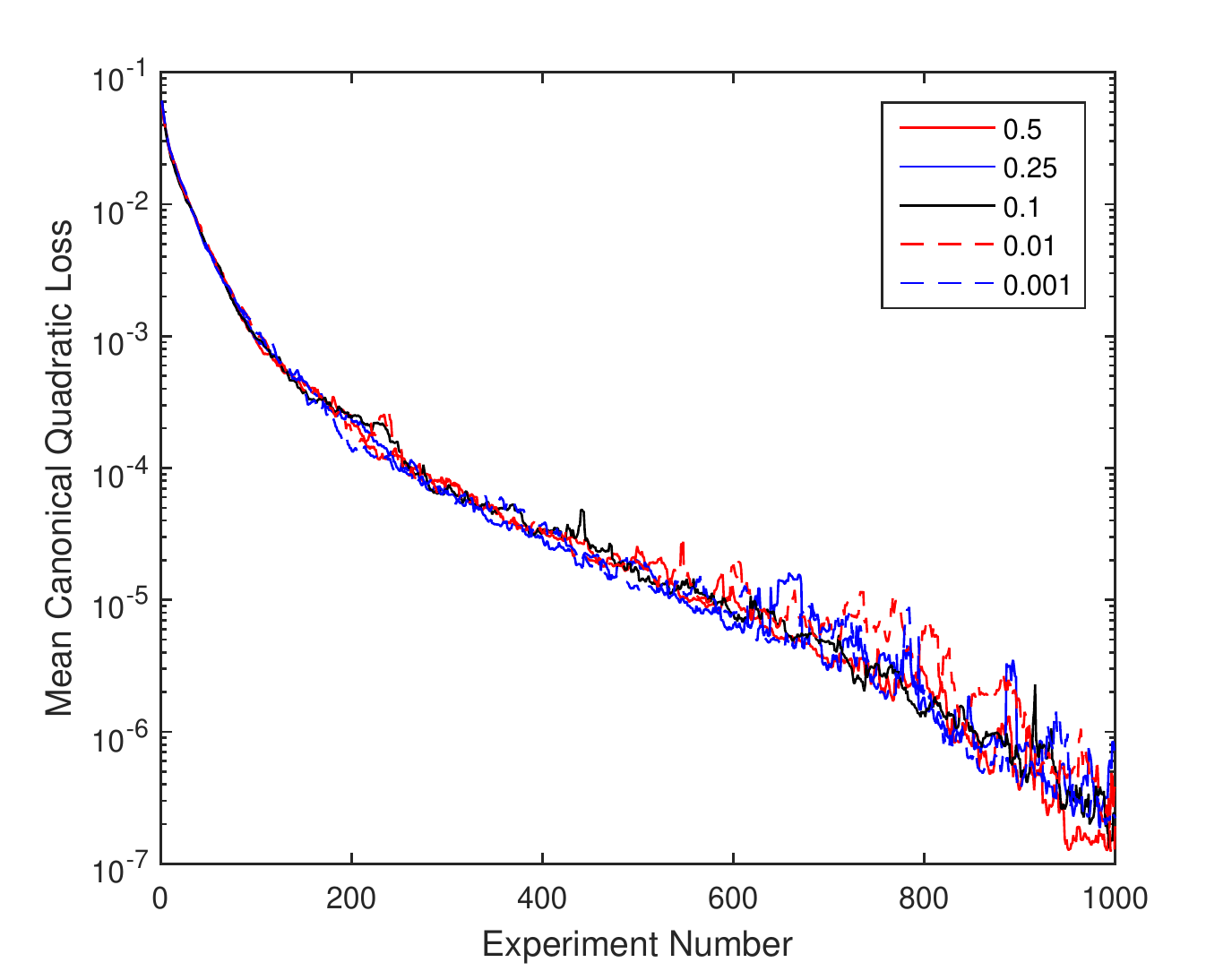}
    \end{minipage}
    \begin{minipage}{\linewidth}
        \parametertable{
            $a$ & $0.98$ & $d_{\max}$ & $4$ \\
            $N_{\rm part}$ & $8000$ & $n_{\rm clusters}$ & $2$ \\
            $h$ & $0.5$ & $\min n_{\rm part}$ & $1000$ \\
            $n_{\rm meas}$ & $3$ & $K_{\rm champ}$ & $2000$ \\
        }
    \end{minipage}
    \caption{
        Median and mean canonical quadratic loss for RGE with $3$ eigenvalues as a function of the floor weights.  All structured parameters are applied
        at the root context.
        \label{fig:Floor}
    }
\end{figure}

\autoref{fig:Floor} investigates the role of the floor weight rule here.  We see that the data is relatively insensitive to the choice of floor weights.  This implies that usually the champion rule is responsible for the majority of the pruning of the structure graph.

\clearpage
\section{Pseudocode}
\label{apx:pseudocode}

\begin{algorithm}[H]
    \caption{\label{alg:weighted-kmeans}
        Weighted $k$-means unsupervised clustering algorithm.
    }
    \begin{algorithmic}
        \Require
            Particle locations $\kof{\vec{x}}{i}{\npart}$,
            particle weights $\kof{\vec{x}}{i}{\npart}$,
            number of clusters $k$.
        \Ensure Cluster labels $\{\ell_i\} \subseteq \{1, \dots, k\}$,
            cluster centroids $\kof{\vec{\mu}}{j}{k}$.
        
        \Function{WeightedKMeans}{$\{\vec{x}_i\}$, $\{w_i\}$, $k$}
            \seccomment{Initialization.}
            \State Let $\{\mu_j\} \gets$ \Call{KMeans++}{$\{\vec{x}_i\}$}.
                \inlinecomment{Initialize centroids.}
            \State Let $\ell_i \gets \argmin_{j} \|\vec{x}_i - \vec{\mu}_j\|^2$
                for each $i \in \{1, \dots, \npart\}$.
                \inlinecomment{Initialize labels from centroids.}
            \seccomment{Iterative improvement.}
            \For{$i_\iter \in 1 \to n_{\iters}$}
                \linecomment{
                    Note that the next line is where weighted
                    and unweighted $k$-means differ, in that
                    we consider the weights $w_i$.
                }
                \State Let $\vec{\mu}_j \gets
                    \sum_{i~\text{s.t.}~\ell_i = j} w_i \vec{x}_i /
                    \sum_{i~\text{s.t.}~\ell_i = j} w_i$
                    for each $j \in \{1, \dots, k\}$.
                    \inlinecomment{Recompute centroids from previous labels.}
                \State Let $\ell_i \gets \argmin_{j} \|\vec{x}_i - \vec{\mu}_j\|^2$
                    for each $i \in \{1, \dots, \npart\}$.
                    \inlinecomment{Recompute labels from new centroids.}
                \If{no labels changed this iteration}
                    \State \Return $\{\ell_i\}$, $\{\vec{\mu}_j\}$.
                \EndIf
            \EndFor
            \seccomment{Error handling.}
            \State \Raise an error indicating too many iterations
                were used.
            \vskip0.5em
        \EndFunction
    \end{algorithmic}
\end{algorithm}

\begin{algorithm}[H]
    \caption{\label{alg:kmeanspp}
        $k$-means++ procedure \cite{arthur_k-means++:_2007}~for initializing centroid locations.
    }
    \begin{algorithmic}
        \Require
            Particle locations $\kof{\vec{x}}{i}{\npart}$,
            number of clusters $k$.
        \Ensure Initial cluster centroids $\{\vec{\mu}_i : i \in \{1, \dots, k\}\}$.
        
        \Function{KMeans++}{$\{\vec{x}_i\}$, $k$}%
            \seccomment{Initialization.}
            \State Draw $i$ uniformly at random from $\{1, \dots, \npart\}$.
            \State Let $\vec{\mu}_1 = \vec{x}_i$.
            \seccomment{Iteration.}
            \For{$j \in 2 \to k$}
                \State Draw $i$ from $\{1, \dots, \npart\}$ with probability $D^2(i) / \sum_i D^2(i)$,
                    where
                    \begin{align*}
                        D(i) \defeq \min_{j' \in \{1, \dots, j - 1\}} \| \vec{x}_i - \vec{\mu}_{j'} \|.
                    \end{align*}
                \State Let $\vec{\mu}_j = \vec{x}_i$.
            \EndFor
            \seccomment{Finalization.}
            \State \Return $\{\vec{\mu}_j\}$
            \vskip0.5em
        \EndFunction
    \end{algorithmic}
\end{algorithm}

\end{document}